\documentclass[12pt]{article}
\usepackage{amsmath}
\usepackage{graphicx,psfrag,epsf}
\usepackage{enumerate}
\usepackage{natbib}
\usepackage{url} 

\usepackage{verbatim}
\usepackage[width=\textwidth]{caption}
\usepackage{amsmath,amssymb,bm}
\usepackage[]{float}
\usepackage[]{placeins}
\usepackage{flafter}
\usepackage[]{longtable}
\usepackage{csvsimple}
\usepackage{verbatim}
\usepackage{amsopn}
\usepackage{color}
\usepackage{pdflscape}
\usepackage{array}
\usepackage{natbib}
\usepackage{xr}
\usepackage{booktabs}
\makeatletter

\newcommand{\blind}{0}

\newtheorem{thm}{Theorem}
\newtheorem{cor}{Corollary}

\newtheorem{lem}{Lemma}
\newtheorem{hyp}{Assumption}

\newcommand{\indep}{\perp \!\!\! \perp}

\newcommand{\E}{\mathbb{E}}

\makeatletter
\def\widebreve{\mathpalette\wide@breve}
\def\wide@breve#1#2{\sbox\z@{$#1#2$}%
     \mathop{\vbox{\m@th\ialign{##\crcr
\kern0.08em\brevefill#1{0.8\wd\z@}\crcr\noalign{\nointerlineskip}%
                    $\hss#1#2\hss$\crcr}}}\limits}
\def\brevefill#1#2{$\m@th\sbox\tw@{$#1($}%
  \hss\resizebox{#2}{\wd\tw@}{\rotatebox[origin=c]{90}{\upshape(}}\hss$}
\makeatletter

\begin{document}

\def\spacingset#1{\renewcommand{\baselinestretch}%
{#1}\small\normalsize} \spacingset{1}


\if0\blind
{
  \title{\bf Clustering and External Validity in Randomized Controlled Trials}
  \author{Antoine Deeb\thanks{
    We are very grateful to  Xavier D'Haultf\oe uille, Kelsey Jack, Jaime Ramirez-Cuellar, Heather Royer, Dick Startz, Doug Steigerwald, Chris Udry, Gonzalo Vazquez-Bare, members of the UCSB econometrics research group, and participants in the UCSB labor lunch for their helpful comments}\hspace{.2cm}\\
    Development Impact Evaluation, World Bank\\
    and \\
    Cl\'{e}ment de Chaisemartin \\
    Department of Economics, Sciences Po}
  \maketitle
} \fi

\if1\blind
{
  \bigskip
  \bigskip
  \bigskip
  \begin{center}
    {\LARGE\bf Title}
\end{center}
  \medskip
} \fi

\bigskip
\begin{abstract}
The randomization inference literature studying randomized controlled trials (RCTs)
assumes that units' potential outcomes are deterministic. This assumption is unlikely to
hold, as stochastic shocks may take place during the experiment. In this paper, we consider
the case of an RCT with individual-level treatment assignment, and we allow for individual-level and cluster-level (e.g. village-level) shocks. We show
that one can draw inference on the ATE conditional on the realizations of
the cluster-level shocks, using heteroskedasticity-robust standard errors, or on the ATE netted
out of those shocks, using cluster-robust standard errors.
\end{abstract}

\noindent%
\vfill

\newpage
\spacingset{1.8} 
\section{Introduction}

In a randomized controlled trial (RCT), it is well known that one can estimate and draw inference on the average treatment effect, if the potential outcomes of units participating in the experiment are non-stochastic, a commonly-made assumption in the randomization inference literature \citep[see, e.g.,][]{Neyman1923,li2017,abadie2020}. In practice, it is often implausible that units' potential outcomes are fixed. For instance, an agricultural household's investment decisions may be affected by the weather conditions in its village during planting season, or by other stochastic shocks. In a model where units' potential outcomes are not fixed but depend on stochastic shocks, the results in the randomization inference literature still hold, conditional on the realizations of the shocks affecting units' potential outcomes. One can estimate and draw inference on the average treatment effect (ATE) \textit{conditional on the shocks that occurred during the experiment.} This may not be a parameter of interest, as it lacks in external validity. For instance, when evaluating the effect of a cash grant on farmers' investment decisions, one may want to know the grant's effect independent of the specific shocks that arose during the experiment, rather than the grant's effect given those specific shocks.

%

\medskip
To fix ideas, we describe our paper in the context of the farmers' cash grant RCT example, but our results apply to all experiments where shocks arising at a more aggregated level than the randomization unit can affect the outcome. We assume that the cash grant is randomly assigned to some households within each village. We relax the assumption of deterministic outcomes and allow household-level as well as village-level shocks to affect farmers' potential investment decisions without and with the grant. Finally, we define two estimands of interest: the ATE conditional on the village-level shocks and the ATE netted out of those shocks.

\medskip
We start by showing that researchers can draw inference on the conditional ATE, by regressing farmers' investment on whether they received the cash grant, using the heteroskedasticity-robust variance estimator. This variance estimator is conservative for the variance of the ATE estimator conditional on the village-level shocks. On the other hand, to draw inference on the unconditional ATE, researchers need to cluster their standard errors at the village level. Indeed, we show that the village-clustered variance estimator is conservative for the unconditional variance of the ATE estimator. As is well-known, this estimator can still lead to over-reject if the number of clusters is too low, so it should only be used with a sufficiently large number of clusters \citep[see Section 4.3 of][for a recent review of cluster-robust inference methods with few clusters]{mackinnon2022cluster}. Moreover, accounting for the village-level shocks increases the variance of the ATE estimator, which will often decrease power. Therefore, when one rejects the null of no effect with the heteroskedasticity-robust variance estimator but not with the cluster-robust one, that may mean that the treatment had an effect given the specific shocks that arose during the experiment but that this conclusion would not generalize under different shocks. But that may also mean that power is too low to detect the unconditional ATE. Interestingly, we also show that clustering at the village level does not always lead to power losses: owing to the conservative nature of both variance estimators, the expectation of the heteroskedasticity-robust estimator may sometimes be larger than the expectation of the clustered one, when the treatment effect is more heterogeneous across farmers than across villages. In such cases, clustering may actually increase power.\medskip 

In a survey of The American Economic Journal: Applied Economics from 2014 to 2016, we found that only 1 out of the 26 published RCTs clustered their standard errors at a level higher than the unit-of-randomization. Therefore, our results provide an easy to implement and often overlooked solution for researchers to assess the external validity of their findings. By external validity, we mean whether results can be extrapolated beyond the specific circumstances that occurred during the experiment. Whether results can be extrapolated to a different population than the one that participated in the experiment is a different question. Recent articles that consider treatment effects extrapolation outside of the estimation sample include, e.g., \cite{dehejia2019local} or \cite{bo2019assessing}.\medskip

To choose at which level to cluster, one first needs to think of which shocks are likely to arise during the experiment. Shocks are post-randomization events that affect the outcome. For example, in the context of the cash-grant RCT, weather events arising after the randomization are shocks. On the other hand, villages' demographic characteristics may affect the outcome, but they are pre-determined, so they are not shocks. In the context of a nationwide job-placement experiment, a post-randomization event affecting the labor market is a shock. Second, one needs to think of the level at which shocks operate. In the cash-grant RCT example, some weather shocks may arise at the village level and may be independent across villages, while other weather shocks may arise at a more aggregated level. In the job-placement experiment example, some labor market shocks arise at the local level (e.g.: a plant closure), while others arise at the national level (e.g.: a change in the Central Bank's interest rate). There could also be some industry-specific shocks, and other shocks affecting all industries, so shocks need not operate at a geographic level. Finally, one needs to cluster at a level where many shocks are likely to operate, while still having sufficiently many clusters to draw valid inference. In the job placement experiment, clustering at a local (e.g. city or regional) level will account for all the shocks taking place at that level, but it will not account for macro-level shocks.\footnote{In that example, one may want to account both for local- and industry-level shocks. We conjecture that doing so may be feasible using a multi-way clustering method \citep[see][]{cameron2012robust,menzel2018bootstrap,davezies2019empirical}, but showing it goes beyond the scope of this paper.} Clustering can only account for shocks arising at a more disaggregated level than the level at which the experiment took place.\footnote{ \cite{hahn2020estimation} propose a framework for dealing with macro shocks in the context of  structural MLE models, a setting that differs from ours.} The above exercise is not a mechanical one: it leads to concrete, context-specific, recommendations on the level at which one should cluster. Importantly, to avoid specification searching, the level of clustering should be pre-specified.\medskip

Our paper shows that the model-based \citep{cameron2015practitioner,wooldridge2003cluster} and design-based \citep{murray1998design,donner2000design,abadie2017} approaches to clustering are not incompatible, and may be fruitfully combined. In RCTs without clustering in the treatment assignment and where the experimental units are not sampled from a larger population, \cite{abadie2017} have argued that clustering standard errors is not needed. Our results lead to a different recommendation: we consider the very same RCTs (with individual-level treatment assignment, and in a finite population of units that are not sampled from a larger population), and argue that if there are cluster-level shocks affecting the potential outcomes, one may want to cluster if one wants to draw inference on the average treatment effect netted out of the shocks. This difference arises because in \cite{abadie2017}, assignment to treatment is the only source of randomness when experimental units are not drawn from a larger population. In contrast, our setup allows for another source of randomness, the cluster-level shocks, that are not under the investigator's control, and that may alter the outcome. There are a number of contexts where such shocks are likely to arise, and we now review two other recent papers that have documented their existence and proposed methods to take them into account.

\medskip
\cite{udrystochastic2019} have also shown that with aggregate shocks, heteroskedasticity-robust variance estimators may understate the true variance of the ATE estimator in an RCT. As one of their applications, they use an RCT in Ghana conducted from 2009 to 2011 where farmers were given a rainfall insurance and a cash grant. They use those treatments as an instrument for agricultural investment, and they show that returns to investment vary with rainfalls. Using their estimated coefficient for the interaction of investment and rainfalls, they compute the distribution of returns to investment under the rainfall distribution observed over the last 65 years in Ghana. They find that the resulting distribution has a much larger variance than the sampling variance from the experiment would suggest. The solution they propose to account for aggregate shocks differs from ours. First, it requires using additional data (e.g. the distribution of rainfalls in the Ghana example) while ours does not. Second, it is designed to account for specific observable shocks (e.g. rainfall shocks in the Ghana example) while ours can account for any type of cluster-level shock, including unobserved ones. Finally, their method can be used to extrapolate the distribution of the treatment effect under a different distribution of shocks than that observed during the experiment. This extrapolation can be made under the assumption that the shocks interact multiplicatively with the treatment effect. The clustering method we propose does not rely on this assumption; accordingly, it can tell us if there is evidence that the cluster-level shocks that arose during the experiment affected the impact of the intervention, but it cannot tell us anything about the intervention's impact under different shocks. Let us illustrate this important difference through an example. Assume an agricultural experiment took place in a rainy year, with some variation in rainfall across regions, but no drought in any region. Clustering at the regional level, the researcher can test if it is still possible to reject the null of no effect, accounting for the variability in the treatment effect induced by rainfall variations from moderate to high. But clustering cannot tell us whether the treatment would have had an effect during a drought year. The method proposed by \cite{udrystochastic2019} can achieve that, under some assumptions.\medskip



\cite{riddell2019interpreting} have also highlighted an issue similar to that we discuss here. By revisiting the results of the Self-Sufficiency Project, they find that post-randomization events can threaten the validity of experimental designs. They give the following example. In a randomized trial of a chemotherapy treatment conducted at one site only, if an outbreak of C-difficile  occurs during treatment, treatment group members will be more likely to die from the outbreak than the control group members due to a weakened immune system. Then without more sites, we can only draw inference on the treatment effect  conditional on the occurrence of a C-difficile outbreak, which is not necessarily the parameter of interest.
 \cite{riddell2019interpreting}
mention that multiple sites may help researchers to
interpret experimental evidence because different post-randomization events may occur in different sites. Our results support that statement, and show that by clustering at the level at which these post-randomization events take place, one can draw inference on the ATE net of these events. \medskip

Finally, many other papers have departed from the randomization inference literature, and have allowed potential outcomes to be stochastic in RCTs \citep[see, e.g.][]{bugni/canay/shaikh:18,bugni/canay/Shaikh:19}.
However, those papers usually assume that potential outcomes are i.i.d. Instead, we consider the case where units' potential outcomes are correlated due to cluster-level shocks.\medskip

We use our results to revisit \cite{karlan2014agricultural}, who study the effects of a rainfall insurance and of a cash grant treatment on farmers' investment decisions. That paper was also revisited by \cite{udrystochastic2019}, who argue that in this context, regional-level weather shocks need to be accounted for. To do so, we cluster standard errors at the regional level at which \cite{udrystochastic2019} argue that weather shocks occur. Doing so, we do not find very different results from those \cite{karlan2014agricultural} had obtained using heteroskedasticity-robust standard errors, thus showing that their results are robust to accounting for the aggregate shocks that arose during the experiment.\medskip

We also revisit \cite{cole2013barriers}, who study the effects of various treatments on farmers' adoption of a rainfall insurance. Using heteroskedasticity-robust standard errors, the authors found that two of their treatments significantly increased adoption. This experiment took place in 37 villages of two districts of the state of Andra Pradesh in India, so the most aggregated level we can cluster at is the village one. Even clustering at this fairly disaggregated level, we find that only one of the two treatments still has a significant effect on adoption. The effect of the second treatment may have been due to the specific village-level shocks that arose during the experiment, and may not replicate under different circumstances. It could also be the case that accounting for the village-level shocks increases the variance of the treatment effect estimator substantially, and reduces power. At any rate, one cannot assert that this treatment would have had an effect under different shocks.\medskip

The take-aways of our paper for applied researchers are as follows. When one rejects the null of no effect without clustering but not with clustering, one can assert that the treatment had an effect, given the specific shocks that arose during the experiment, but one cannot assert that this conclusion would generalize under different shocks. This may either be because the unconditional ATE is closer to 0 than the conditional ATE, or because accounting for the shocks reduces the study's power and precludes it from detecting the unconditional ATE. On the other hand, when one rejects the null of no effect with clustering, one can assert that the treatment had an effect, independent of the specific shocks that arose during the experiment. Then, the decision to cluster or not depends on the level of external validity one would like to achieve.\medskip

\section{Setup and finite-sample results}

\subsection{Setup and Notation}

We consider an RCT taking place in a finite population of $K$ villages. Village $k$ has $n_k$ households, and randomization is stratified at the village level. The experiment wants to look at the effect of cash grants on farming households' investment in agriculture. The outcomes of interest are households' investments in agriculture such as land preparation costs, value of chemicals used, and acres cultivated. It is arguably implausible to assume that households' potential outcomes are fixed, they may be affected by a wealth of stochastic events that could take place  until the time they make their investment decisions. These shocks could be specific to the households (such as the breadwinner being laid off or injured), or they could be common to all households within a village (such as extreme weather events, economic hardships in the village etc.). We therefore assume that for all $(i,k)\in\{1,...,n_k\}\times\{1,...,K\}$, the potential outcomes of household $i$ in village $k$ without and with the treatment, $Y_{ik}(0)$ and $Y_{ik}(1)$ satisfy the following equations:
\begin{hyp}\label{hyp:outcomes}  Stochastic Potential Outcomes
\begin{eqnarray}\label{eq:Model-Cluster}
Y_{ik}(0)&=& y_{ik}(0) +\eta_k(0) + \epsilon_{ik}(0) \nonumber\\
Y_{ik}(1)&=& y_{ik}(1) + \eta_k(1) + \epsilon_{ik}(1).
\end{eqnarray}
\end{hyp}
$\epsilon_{ik}(0)$ (resp. $\epsilon_{ik}(1)$) represents a shock affecting the potential outcomes of household $i$ in village $k$ if she is untreated (resp. treated). $\eta_{k}(0)$ (resp. $\eta_{k}(1)$) represents a shock affecting all the untreated (resp. treated) households in village $k$. We assume that $\E(\epsilon_{ik}(0))=\E(\epsilon_{ik}(1))=\E(\eta_{k}(0))=\E(\eta_{k}(1))=0$, so $y_{ik}(0)$ (resp. $y_{ik}(0)$) represent the expectation of $Y_{ik}(0)$ (resp. $Y_{ik}(1)$), the outcome without (resp. with) treatment that household $i$ in village $k$ will obtain under ``average'' household- and village-level shocks. In our cash-grant example, $y_{ik}(d)$ is a household's investment under average shocks and treatment $d$. $\epsilon_{ik}(d)$ represents the effect of household-level shocks, such the breadwinner being laid off, on the household's investment under treatment $d$. $\eta_k(d)$ represents the effect of  village level shocks, such as an extreme weather event, on the household's investment under treatment $d$. Let $(\boldsymbol{\eta(0)},\boldsymbol{\eta(1)})=(\eta_{k}(0),\eta_{k}(1))_{1\leq k\leq K}$ be a vector stacking all the village-level shocks, and let $(\boldsymbol{\epsilon(0)},\boldsymbol{\epsilon(1)})=(\epsilon_{ik}(0),\epsilon_{ik}(1))_{1\leq i\leq n_k,1\leq k\leq K}$ be a vector stacking all the household-level shocks.\medskip

Assumption \ref{hyp:outcomes} requires that the shocks be additively separable, and take place at the level of the experimental strata. These two conditions are not of essence for our results to hold. Our main results still hold if shocks take place at a more aggregated level than the experimental strata, provided that the shocks and the treatments remain independent across clusters. Our main results also still hold if the shocks do not affect the potential outcomes in an additively separable manner, i.e. if $Y_{ik}(d)=f_{ikd}\left(\epsilon_{ik}(d),\eta_k(d)\right)$ for some functions $f_{ikd}(.)$. In that case, one just needs to redefine $ATE(\boldsymbol{\eta}(0), \boldsymbol{\eta}(1))$ below as $\frac{1}{n}\sum_{i,k}E(Y_{ik}(1)-Y_{ik}(0)|(\boldsymbol{\eta}(0), \boldsymbol{\eta}(1)))$, and $ATE$ as $\frac{1}{n}\sum_{i,k}E(Y_{ik}(1)-Y_{ik}(0))$, see Section \ref{section-noassum1} of the Appendix for more details. We expect most readers to be familiar with the additively separable model, so we stick to it in the paper to facilitate reading.\medskip

Let $n=\sum_{k=1}^Kn_k$ denote the total number of households in the $K$ villages. We may be more interested in learning
\begin{equation}
ATE= \frac{1}{n}\sum_{i,k}[y_{ik}(1)-y_{ik}(0)],
\end{equation}
rather than
\begin{equation}
ATE(\boldsymbol{\epsilon}(0), \boldsymbol{\epsilon}(1),\boldsymbol{\eta(0)},\boldsymbol{\eta(1)}) =\frac{1}{n}\sum_{i,k}[Y_{ik}(1)-Y_{ik}(0)],
\end{equation}
or
\begin{eqnarray}
ATE(\boldsymbol{\eta}(0), \boldsymbol{\eta}(1))    = \frac{1}{n}\sum_{i,k}\left[(y_{ik}(1)+\eta_k(1))-(y_{ik}(0)+\eta_k(0))\right].
\end{eqnarray}

The second parameter is the average effect of the treatment on households' investments, conditional on the specific village and household shocks that arose during the experiment. The third parameter is the average effect of the treatment on households' investments, conditional only on the specific village shocks. The first parameter is the average effect of the treatment, net of those specific shocks. This parameter is more externally valid than the other two, as it applies beyond the specific circumstances that occurred during the experiment.\footnote{On the other hand, $ATE$ still only applies to the villages participating in the experiment.}
\medskip

Let $D_{ik}$ be an indicator for whether household $i$ in village $k$ is treated, let $\textbf{D}_k$  be a vector stacking the treatment indicators of all households in village $k$, and let $\textbf{D}$ be a matrix stacking these vectors. We consider the following assumption:

\begin{hyp}\label{hyp:main_general}
For all $i,k$,
\begin{enumerate}
\item  $V(\epsilon_{ik}(d))=\sigma_{dik}^2<+\infty$ for $d=0,1$. \label{hyp-point:finite-var-eps}
\item For all $j\ne i$, $(\epsilon_{ik}(0),\epsilon_{ik}(1))\indep (\epsilon_{jk}(0),\epsilon_{jk}(1))$. \label{hyp-point:indep-among-eps}
\item  $\textbf{D} \indep \left(\left(\epsilon_{ik}(0),\epsilon_{ik}(1)\right)_{1\leq i \leq n_k},\eta_k(0),\eta_k(1)\right)_{1\leq k \leq K}$. \label{hyp-point:treatment-indep}
\item  $(\epsilon_{ik}(0),\epsilon_{ik}(1))_{1\leq i \leq n_k}\indep (\eta_{k}(0),\eta_{k}(1))_{1\leq k \leq K}$. \label{hyp-point:indep-eps-eta}
\item $V(\eta_{k}(1)-\eta_{k}(0))<+\infty$. \label{hyp-point:finite-var-etas}
\end{enumerate}
\end{hyp}

Point \ref{hyp-point:finite-var-eps} requires that  $(\epsilon_{ik}(0), \epsilon_{ik}(1))$ have a second moment. Point \ref{hyp-point:indep-among-eps} requires that in each village, the household level shocks be independent. Point \ref{hyp-point:treatment-indep} requires that the household- and village-level shocks be independent of the treatments, which usually holds by design in a RCT. Point \ref{hyp-point:indep-eps-eta} requires that the household- and village-level shocks be independent. Finally, Point \ref{hyp-point:finite-var-etas} requires that the variance of $\eta_{k}(1)-\eta_{k}(0)$ exist. Assumption \ref{hyp:main_general} does not require that the shocks $(\epsilon_{ik}(0),\epsilon_{ik}(1))$ and $(\eta_{k}(0),\eta_{k}(1))$ be identically distributed: the variance of the shocks may for instance vary across households or villages. Assumption \ref{hyp:main_general} also does not require that $\epsilon_{ik}(0)$ and $\epsilon_{ik}(1)$  be independent, or that $\eta_{k}(0)$ and $\eta_{k}(1)$ be independent: one may for instance have $\epsilon_{ik}(0)=\epsilon_{ik}(1)$ and $\eta_{k}(0)=\eta_{k}(1)$, if the household- and village-level shocks are the same when treated and untreated. \medskip

Let $n_{1k}$ and $n_{0k}$ respectively denote the number of households in the treatment and control groups in village $k$. Let $Y_{ik}=D_{ik}Y_{ik}(1)+(1-D_{ik})Y_{ik}(0)$ denote the observed outcome of household $i$. For any variable $x_{ik}$ defined for every $i\in\{1,...,n_k\}$ and $k\in\{1,...,K\}$, let $\overline{x}_k=\frac{1}{n_k}\sum_{i=1}^{n_k}x_{ik}$ denote the average value of $x_{ik}$ in village $k$,  let $\overline{x}_{1k}=\frac{1}{n_{1k}}\sum_{i=1}^{n_{1k}}D_{ik}x_{ik}$ and $\overline{x}_{0k}=\frac{1}{n_{0k}}\sum_{i=1}^{n_{0k}}(1-D_{ik})x_{ik}$ respectively denote the average value of $x_{ik}$ among the treated and untreated households in village $k$, and let $\overline{x}=\frac{1}{n}\sum_{i,k}x_{ik}$ denote the average value of $x_{ik}$ across all households.

Then let $\overline{n}=\frac{n}{K}$, and let

\begin{eqnarray*}
\widehat{ATE}_k     &=& \overline{Y}_{1k}-\overline{Y}_{0k}\\
\widehat{ATE}      &=&\frac{1}{K}\underset{k=1}{\overset{K}{\sum}}\frac{n_k}{\overline{n}}\widehat{ATE}_k,
\end{eqnarray*}

respectively denote the standard difference in means estimator of the average treatment effect in village $k$, and the estimated average treatment effect in the $K$ villages.

For any variable $x_{ik}$ defined for every $i\in\{1,...,n_k\}$ and $k\in\{1,...,K\}$, let $S^2_{x,k}=\frac{1}{n_k-1}\sum_{i=1}^{n_k}(x_{ik}-\overline{x}_k)^2$ denote the variance of $x_{ik}$ in village $k$, and let $S^2_{x,1,k}=\frac{1}{n_{1k}-1}\sum_{i=1}^{n_{1k}}D_{ik}(x_{ik}-\overline{x}_{1k})^2$ and $S^2_{x,0,k}=\frac{1}{n_{0k}-1}\sum_{i=1}^{n_{0k}}(1-D_{ik})(x_{ik}-\overline{x}_{0k})^2$ respectively denote the variance of $x_{ik}$ among the treated and untreated households in village $k$.
Then let,
\begin{eqnarray*}
\widehat{V}_{rob}\left(\widehat{ATE}_k\right)&=& \frac{1}{n_{1k}}  S^2_{Y,1,k}+\frac{1}{n_{0k}} S^2_{Y,0,k} \\
\end{eqnarray*}
denote the robust estimator of the variance of $\widehat{ATE}_k$ \citep{eicker1963asymptotic,huber1967behavior,white1980heteroskedasticity}, and let
\begin{eqnarray*}
\widehat{V}_{rob}\left(\widehat{ATE}\right)&=&\frac{1}{K^2}\underset{k=1}{\overset{K}{\sum}}\left(\frac{n_k}{\overline{n}}\right)^2 \widehat{V}_{rob}\left(\widehat{ATE}_k\right),
\end{eqnarray*}
denote the estimator of the variance of $\widehat{ATE}$ one can form using those estimators and assuming the $\widehat{ATE}_k$s are independent.

\medskip
We assume that the treatment is randomly assigned at the household level in each village:
\begin{hyp}\label{hyp:strat_completely_randomized}
Stratified completely randomized experiment \\
\textcolor{black}{For all $k$, $\sum_{i=1}^{n_k}D_{ik}=n_{1k}$, and for every $(d_1,...,d_{n_k})$ such that $d_{1}+...+d_{n_k}=n_{1k}$}, $P(D_k=(d_1,...,d_{n_k}))=\frac{1}{{n_k \choose n_{1k}}}$.
\end{hyp}
Finally, we make the following assumption:
\begin{hyp}\label{hyp:mutual-indep}
The vectors $\left(\textbf{D}_k,\eta_k(1),\eta_k(0),(\epsilon_{ik}(0),\epsilon_{ik}(1))_{1\leq i \leq n_k}\right)$ are mutually independent.
\end{hyp}

Assumption \ref{hyp:mutual-indep} requires that the variables attached to different villages be mutually independent. 

%

\subsection{Finite-sample results}

We can now state our first result.

\begin{thm}\label{thm:cluster-conditional}
If Assumptions  \ref{hyp:outcomes}, \ref{hyp:main_general}, \ref{hyp:strat_completely_randomized}, and \ref{hyp:mutual-indep}  hold,
\begin{align*}
1.\ \E\left(\widehat{ATE}\middle|\boldsymbol{\eta}(0), \boldsymbol{\eta}(1)\right)&=ATE(\boldsymbol{\eta}(0), \boldsymbol{\eta}(1)) \\
2. \ V\left(\widehat{ATE}\middle|\boldsymbol{\eta}(0), \boldsymbol{\eta}(1)\right)&=\frac{1}{K^2}\underset{k=1}{\overset{K}{\sum}}\left(\frac{n_k}{\overline{n}}\right)^2V\left(\widehat{ATE}_k\middle|\boldsymbol{\eta}(0), \boldsymbol{\eta}(1)\right), \text{where}\\
\noalign{\ \ \  \ \ \ \ $V\left(\widehat{ATE}_k\middle|\boldsymbol{\eta}(0), \boldsymbol{\eta}(1)\right) = \frac{1}{n_{0k}} S^2_{y(0),k} + \frac{1}{n_{1k}} S^2_{y(1),k} - \frac{1}{n_k} S^2_{y(1)-y(0),k}+\frac{1}{n_{1k}}\overline{\sigma^2_1}_k+  \frac{1}{n_{0k}}\overline{\sigma^2_0}_k$}
3. \ V\left(\widehat{ATE}\middle|\boldsymbol{\eta}(0), \boldsymbol{\eta}(1)\right)  &\leq \E\left(\widehat{V}_{rob}(\widehat{ATE})\middle|\boldsymbol{\eta}(0), \boldsymbol{\eta}(1)\right), \text{with equality if }\\
\noalign{ \ \ \  \ \ \ \ \text{there is no treatment effect heterogeneity within village: $S^2_{y(1)-y(0),k}=0$ for all $k$.}}
\end{align*}
\end{thm}
Point 1 of Theorem \ref{thm:cluster-conditional} shows that $\widehat{ATE}$ is an unbiased estimator of $ATE(\boldsymbol{\eta}(0), \boldsymbol{\eta}(1))$, conditional on the village-level shocks. Point 2 gives a formula for the variance of $\widehat{ATE}$ conditional on the village-level shocks. It is similar to the variance of $\widehat{ATE}$ in \citet{Neyman1923}, derived assuming fixed potential outcomes. However, it contains one more term, $\frac{1}{K^2}\underset{k=1}{\overset{K}{\sum}}\frac{1}{n_{1k}}\overline{\sigma^2_1}_k+  \frac{1}{n_{0k}}\overline{\sigma^2_0}_k$, which comes from the added variation created by the individual-level shocks. Point 3 shows that the robust variance estimator is a conservative estimator of that conditional variance.\medskip

In our set-up, the result in \cite{Neyman1923} implies that conditional on the household- and village-level shocks, $\widehat{ATE}$ is an unbiased estimator, and the robust variance estimator is a conservative estimator of the variance of $\widehat{ATE}$. Theorem \ref{thm:cluster-conditional} extends this result, by showing that it still holds when one only conditions on the village-level shocks.\footnote{It has also been shown  \citep[see][]{imbens2015} that when the potential outcomes are i.i.d., the robust variance estimator is an unbiased estimator of  $V(\widehat{ATE})$. This result can also be obtained from Theorem  \ref{thm:cluster-conditional}. Assume that $\eta_k(0)=\eta_k(1)=0$, thus ensuring that the potential outcomes are independent, and that $y_{ik}(d)=y(d)$ and $\sigma^2_{dik}=\sigma^2_{d}$, thus ensuring that they are identically distributed. Then, Point 3 implies that $V\left(\widehat{ATE}\right)= \E\left(\widehat{V}_{rob}\left(\widehat{ATE}\right)\right)$.} Conservative variance estimators are not specific to this paper and are commonly found in the randomization inference literature considering RCT samples as a fixed population rather than a random draw from a super-population \citep[see, e.g.,][]{abadie2020}.\medskip

In our cash-grant example, Theorem \ref{thm:cluster-conditional} implies that if the researcher uses robust standard errors and finds a statistically significant effect, she can conclude that $ATE(\boldsymbol{\eta}(0), \boldsymbol{\eta}(1)) \ne 0$: the treatment had an effect, given the specific village-level shocks that arose during the experiment. $ATE(\boldsymbol{\eta}(0),\boldsymbol{\eta}(1))$ does not depend on the household-level shocks that arose during the experiment, but it does depend on the village-level shocks. Therefore, the researcher cannot say whether the treatment would still have had an effect if different village-level shocks had occurred. To answer that question, one needs to draw inference on $ATE$. We now show that this can be achieved, by clustering standard errors at the village level.
Let
\begin{equation*}
\widehat{V}_{clu}(\widehat{ATE})= \frac{1}{K\left(K-1\right)}\sum_{k=1}^K \left(\frac{n_k}{\overline{n}}\widehat{ATE}_k-\widehat{ATE}\right)^2
\end{equation*}
be the cluster-robust estimator of the variance of $\widehat{ATE}$ \citep{liang1986longitudinal}. Specifically, up to a degrees of freedom adjustment, $\widehat{V}_{clu}(\widehat{ATE})$ is equal to the cluster-robust estimator of the variance of the treatment coefficient in a regression of the outcome on a constant and the treatment, clustered at the strata level, with propensity score reweighting to account for the fact that treatment probabilities vary across clusters.

\begin{thm}\label{thm:main-cluster}

If Assumptions \ref{hyp:outcomes}, \ref{hyp:main_general}, \ref{hyp:strat_completely_randomized}, and \ref{hyp:mutual-indep} hold,
\begin{align*}
1. \ &\E\left(\widehat{ATE}\right)=ATE\\
2. \ &V\left(\widehat{ATE}\right)= \frac{1}{K^2}\underset{k=1}{\overset{K}{\sum}}\left(\frac{n_k}{\overline{n}}\right)^2\bigg[\frac{1}{n_{0k}} S^2_{y(0),k} + \frac{1}{n_{1k}} S^2_{y(1),k} - \frac{1}{n_k} S^2_{y(1)-y(0),k}+\frac{1}{n_{1k}}\overline{\sigma^2_1}_k+\\
& \ \ \ \ \frac{1}{n_{0k}}\overline{\sigma^2_0}_k+V\left(\eta_k(1)-\eta_k(0)\right)\bigg].  \\
3. \ &\E\left(V\left(\widehat{ATE}\middle|\boldsymbol{\eta}(0), \boldsymbol{\eta}(1) \right)\right)\leq V\left(\widehat{ATE}\right)\leq \E\left[\widehat{V}_{clu}(\widehat{ATE})\right].\\
\noalign{\text{The second inequality is an equality if $n_k=\overline{n}$ and $ATE_k=ATE$,}}
\noalign{\text{ with $ATE_k=\frac{1}{n_k}\sum_{i=1}^{n_k}\left[y_{ik}(1)-y_{ik}(0)\right]$ for all $k=1,...,K$.}}
\end{align*}
\end{thm}
Point 1 of Theorem \ref{thm:main-cluster} shows that $\widehat{ATE}$ is an unbiased estimator of $ATE$. Point 2 gives a formula for the unconditional variance of $\widehat{ATE}$.  Of course, the unconditional variance of $\widehat{ATE}$ is larger than the conditional one: accounting for the shocks increases the variance, which will often though not always decrease power (see below for further discussion). Point 3 shows that the cluster-robust variance estimator is a conservative estimator of the unconditional and conditional variances of $\widehat{ATE}$. This estimator can still lead to over-reject if the number of clusters is too low, so it should only be used with a sufficiently large number of clusters \citep[see Section 4.3 of][for a recent review of cluster-robust inference methods with few clusters]{mackinnon2022cluster}.\medskip

In our cash-grant example, Theorem \ref{thm:main-cluster} states that if the researcher uses the cluster-robust standard errors and finds a statistically significant effect, she can conclude that $ATE\ne 0$. $ATE$ does not depend on the household- and village-level shocks that arose during the experiment. Therefore, this conclusion is not dependent on the specific shocks that arose during the experiment, but holds when the shocks are averaged out. \medskip

Our approach comes with a risk. By defining several potential estimands of interest, it may lead researchers to test several null hypothesis, with or without clustering, or clustering at various different levels. This would distort inference. To avoid that risk, researchers should pre-commit to an analysis plan that specifies if and at what level they intend to cluster standard errors. \medskip

\begin{cor} \label{cor:vars}

If Assumptions \ref{hyp:outcomes}, \ref{hyp:main_general}, \ref{hyp:strat_completely_randomized}, and \ref{hyp:mutual-indep} hold and $n_k=\overline{n}$ for all $k$,
\begin{align*}
 &\E\left[K\widehat{V}_{clu}(\widehat{ATE})\right]-\E\left[K\widehat{V}_{rob}(\widehat{ATE})\right]\\
 =& \frac{1}{K}\underset{k=1}{\overset{K}{\sum}}V\left(\eta_k(1)-\eta_k(0)\right)
+\frac{1}{K-1}\underset{k=1}{\overset{K}{\sum}}\left(\E\left(\widehat{ATE}_k\right)-\frac{1}{K}\underset{k'=1}{\overset{K}{\sum}}\E\left(\widehat{ATE}_{k'}\right)\right)^2\\ -&\frac{1}{\overline{n}}\frac{1}{K}\underset{k=1}{\overset{K}{\sum}}S^2_{y(1)-y(0),k}
\end{align*}
\end{cor}
Corollary \ref{cor:vars} states that the difference between the expectations of the normalized clustered and robust variance estimators is equal to the average of $V(\eta_k(1)- \eta_k(0))$, a term that comes from the fact the clustered estimator accounts for the shocks, plus the difference between the variance of the treatment effect between villages and the average variance of the treatment effect within villages divided by $\overline{n}$, a term that comes from the fact both variance estimators are conservative. This corollary has two important implications. First, if households' potential outcomes are identically distributed, then $y_{ik}(1)- y_{ik}(0)=\tau$ for all $(i,k)$, and $\E\left[K\widehat{V}_{clu}(\widehat{ATE})\right]-\E\left[K\widehat{V}_{rob}(\widehat{ATE})\right] =\frac{1}{K}\underset{k=1}{\overset{K}{\sum}}V\left(\eta_k(1)-\eta_k(0)\right)$. Therefore, one can test whether there are village-level shocks that affect the impact of the intervention by testing whether the two variance estimators significantly differ.

\medskip
Second, consider the following assumption:
\begin{hyp}\label{hyp:homog-shocks}
Homogeneous clustered shocks\\
For all $k$, $\eta_k(1)=\eta_k(0)$.
\end{hyp}
Assumption \ref{hyp:homog-shocks} requires that treated and untreated households are affected similarly by the village-level shocks.\footnote{It is not testable without imposing other assumptions. Under the assumption that $ATE_k$ does not vary across $k$, one can test whether the $\widehat{ATE}_k$s significantly differ. If they do, that implies that $\eta_k(1)\ne\eta_k(0)$ for some $k$.}
Under Assumptions \ref{hyp:outcomes} and \ref{hyp:homog-shocks},  $ATE(\boldsymbol{\eta}(0), \boldsymbol{\eta}(1))=ATE$ and  $V\left(\widehat{ATE}\right)=V\left(\widehat{ATE}|\boldsymbol{\eta}(0), \boldsymbol{\eta}(1)\right)$, so Theorems \ref{thm:cluster-conditional} and \ref{thm:main-cluster} imply that both $\widehat{V}_{rob}(\widehat{ATE})$ and $\widehat{V}_{clu}(\widehat{ATE})$ are conservative for $V\left(\widehat{ATE}\right)$. Corollary \ref{cor:vars} shows that $\widehat{V}_{clu}(\widehat{ATE})$ can be less conservative than $\widehat{V}_{rob}(\widehat{ATE})$, if there is
more treatment effect heterogeneity within rather than between villages. When Assumption \ref{hyp:homog-shocks} fails, Theorems \ref{thm:cluster-conditional} and \ref{thm:main-cluster} imply that both $\widehat{V}_{rob}(\widehat{ATE})$ and $\widehat{V}_{clu}(\widehat{ATE})$ are conservative for $\E\left(V\left(\widehat{ATE}\middle|\boldsymbol{\eta}(0), \boldsymbol{\eta}(1)\right)\right)$, and Corollary \ref{cor:vars} shows that $\widehat{V}_{clu}(\widehat{ATE})$ can be less conservative than $\widehat{V}_{rob}(\widehat{ATE})$ if the additional variance in $\widehat{ATE}$ coming from the village-level shocks is lower than the difference between the within-village variance of the treatment effect divided by $\overline{n}$ and the between-village variance of the treatment effect. Thus, clustering may not always reduce power.

\section{Large-sample results}

We now derive the asymptotic distribution of $\widehat{ATE}$ considering a case where the number of villages $K$ goes to infinity. First let:
\begin{eqnarray*}
AD_k&=&\frac{n_k}{\bar{n}}\widehat{ATE}_k,
\end{eqnarray*}

and consider the following assumption:

\begin{hyp}\label{hyp:reg-cond} Regularity conditions to derive the asymptotic distribution of $\widehat{ATE}$\\
For some $\epsilon>0$,
\begin{enumerate}
\item For every $k$, $\E\left(AD_k^{2+\epsilon}\right)\leq M < +\infty$, for some $M>0$.
\item  $\underset{K\rightarrow +\infty}{\lim}\frac{1}{S_K^{2+\epsilon}}\sum_{k=1}^{K}\E\left[|AD_k-E(AD_k)|^{2+\epsilon}\right]=0$, where $S_K^2=\sum_{k=1}^{K} V(AD_k)$.
\item $\frac{1}{K}\sum_{k=1}^K\E\left(AD_k\right)$, $\frac{1}{K}\sum_{k=1}^K\E\left(AD_k^2\right)$, and $\frac{1}{K}\sum_{k=1}^{K}\E\left(AD_k\right)^2$ converge towards finite limits when $K\rightarrow \infty$.
\end{enumerate}
\end{hyp}

Assumption  \ref{hyp:reg-cond} contains the regularity conditions needed to apply the strong law of large numbers in  Lemma 1 of \cite{liu1988bootstrap} and the Lyapunov CLT. Also let:

\begin{eqnarray*}
\sigma^2&=& \lim_{K\rightarrow\infty} \frac{1}{K}\sum_{k=1}^K \E\left(AD_k^2\right) -\frac{1}{K}\sum_{k=1}^{K}\E\left(AD_k\right)^2,\\
\sigma_+^2&=&  \lim_{K\rightarrow\infty} \frac{1}{K}\sum_{k=1}^K \E\left(AD_k^2\right) -\left(\frac{1}{K}\sum_{k=1}^{K}\E\left(AD_k\right)\right)^2.
\end{eqnarray*}

We show:
\begin{thm}\label{thm:asym-uncon}
If Assumptions \ref{hyp:outcomes}, \ref{hyp:main_general}, \ref{hyp:strat_completely_randomized}, \ref{hyp:mutual-indep}, and \ref{hyp:reg-cond} hold,
\begin{eqnarray*}
&1.& \ \sqrt{K}\left(\widehat{ATE}-ATE\right)\overset{d}{\rightarrow} N\left(0,\sigma^2\right).\\
&2.& \ K\widehat{V}_{clu}\left(\widehat{ATE}\right) \overset{p}{\rightarrow} \sigma_+^2 \geq \sigma^2.
\end{eqnarray*}
\end{thm}

Point 1 of Theorem \ref{thm:asym-uncon} shows that $\widehat{ATE}$ is an asymptotically normal estimator of $ATE$ when the number of villages goes to infinity. Point 2 shows that $K\widehat{V}_{clu}\left(\widehat{ATE}\right)$  converges to a finite  upper bound of the asymptotic variance of $\widehat{ATE}$ and can be used to construct conservative confidence intervals for $ATE$.

\section{Applications}
\subsection{Agricultural Decisions After Relaxing Credit And Risk Constraints \label{section:karlan}}

Table 4 in \cite{karlan2014agricultural} presents the effects of having rainfall index insurance, receiving a capital grant, and having both treatments on  investment decisions and value of harvest. The results are obtained using the first two years of a three-year RCT conducted in Ghana. In the first year, the authors randomly assigned households to one of four groups: the cash grant group, the insurance group, the cash grant and insurance group, and the control group. In the second year, the cash grant experiment was still present but the insurance grant experiment was replaced by an insurance pricing experiment. Insurance prices were randomized at the community level but every community also had control households without access to the insurance with the randomization being at the household level. For farmers that were offered insurance, insurance take-up is instrumented using the price offered to them \citep[see][]{karlan2014agricultural}.\medskip

In a re-analysis of this experiment, \cite{udrystochastic2019} divide communities into 11 regions, and show that returns to farmers' investments respond to the weather shocks affecting their region. In Table \ref{table:appli} below, Panel A replicates the results in \cite{karlan2014agricultural}, using heteroskedasticity-robust variance estimators. In Panel B, we instead use cluster-robust variance estimators, clustering at the level of the 11 regions indicated by \cite{udrystochastic2019}. As there are only 11 clusters, in Panel C we present p-values computed using the wild-bootstrap test proposed in \cite{cameron2008bootstrap}, and that has been shown to have good properties with a small number of large clusters, see \cite{canay2019wild}. Clustering at the region level does not strongly affect the results in \cite{karlan2014agricultural}. The only exception is for the outcome ``value of chemicals used'', for which treatment effects are less significant with the wild cluster bootstrap than with robust standard errors. Otherwise, for most of the outcomes for which we can reject $ATE(\eta(0),\eta(1))=0$, we can also reject $ATE=0$.
\begin{table}[H]
\caption{Effects in \cite{karlan2014agricultural}, without and with clustering. \label{table:appli}}
	\resizebox{\textwidth}{!}{%
\begin{tabular}{l*{8}{c}}
\toprule
            &\multicolumn{1}{c}{(1)}&\multicolumn{1}{c}{(2)}&\multicolumn{1}{c}{(3)}&\multicolumn{1}{c}{(4)}&\multicolumn{1}{c}{(5)}&\multicolumn{1}{c}{(6)}&\multicolumn{1}{c}{(7)}\\
            &\multicolumn{1}{c}{ Land  }&\multicolumn{1}{c}{\# of Acres}&\multicolumn{1}{c}{  Value of }&\multicolumn{1}{c}{ Wages Paid  }&\multicolumn{1}{c}{  Opportunity Cost   }&\multicolumn{1}{c}{ Total }&\multicolumn{1}{c}{Value of }\\
            &\multicolumn{1}{c}{Preparation Costs }&\multicolumn{1}{c}{Cultivated}&\multicolumn{1}{c}{  Chemicals Used }&\multicolumn{1}{c}{ to Hired Labor }&\multicolumn{1}{c}{ of Family  Labor }&\multicolumn{1}{c}{Costs  }&\multicolumn{1}{c}{ Harvest  }\\
\midrule
\multicolumn{1}{l}{\textbf{A: Robust SE  }} \\
\addlinespace
Insured&      25.528** &       1.024** &      37.904** &      83.537   &      98.161   &     266.146** &     104.274   \\
            &    (12.064)   &     (0.420)   &    (14.854)   &    (59.623)   &    (84.349)   &   (134.229)   &    (81.198)   \\
             &     [0.034]   &     [0.015]   &     [0.011]   &     [0.161]   &     [0.245]   &     [0.047]   &     [0.199]   \\
Insured*Capital Grant    &      15.767   &       0.257   &      66.440***&      39.760   &     -52.653   &      72.137   &     129.243   \\
            &    (13.040)   &     (0.445)   &    (15.674)   &    (65.040)   &    (86.100)   &   (138.640)   &    (81.389)   \\
            &     [0.227]   &     [0.563]   &     [0.000]   &     [0.541]   &     [0.541]   &     [0.603]   &     [0.112]   \\
Capital Grant&      15.362   &       0.088   &      55.631***&      75.609   &    -130.562   &       2.438   &      64.822   \\
            &    (13.361)   &     (0.480)   &    (17.274)   &    (68.914)   &    (92.217)   &   (148.553)   &    (89.764)   \\
             &     [0.250]   &     [0.854]   &     [0.001]   &     [0.273]   &     [0.157]   &     [0.987]   &     [0.470]   \\
\midrule
\multicolumn{1}{l}{\textbf{B: Clustered SE }} \\
\addlinespace
Insured&      25.528** &       1.024***&      37.904** &      83.537   &      98.161   &     266.146***&     104.274*  \\
            &    (12.498)   &     (0.372)   &    (17.784)   &    (52.591)   &    (67.068)   &    (97.865)   &    (60.776)   \\
            & [0.041] & [0.006] & [0.033] &[0.112] & [0.143] & [0.007] & [ 0.086] \\

Insured*Capital Grant     &      15.767   &       0.257   &      66.440***&      39.760   &     -52.653   &      72.137   &     129.243*  \\
            &    (14.307)   &     (0.266)   &    (12.018)   &    (53.571)   &    (56.916)   &    (94.551)   &    (74.076)   \\
           &[0.270]& [0.332]& [0.000] & [0.458] & [0.355] & [0.445] &  [0.081] \\

Capital Grant&      15.362   &       0.088   &      55.631** &      75.609   &    -130.562   &       2.438   &      64.822   \\
            &    (15.092)   &     (0.504)   &    (25.531)   &    (50.493)   &   (114.161)   &   (185.320)   &   (122.354)   \\
            & [0.309] & [ 0.861] & [0.029] & [0.134] & [0.253]& [0.990] &  [0.596] \\
\midrule
\multicolumn{1}{l}{\textbf{C: Wild Bootstrap  }} \\
\addlinespace
Insured&      25.528* &       1.024**&      37.904 &      83.537   &      98.161   &     266.146*&     104.274  \\
            &    [0.069]   &      [0.032]  &    [0.141]  &    [0.103]   &    [0.299]   &    [0.060]   &    [0.106]   \\
Insured*Capital Grant    &       15.767   &       0.257   &      66.440**&      39.760   &     -52.653   &      72.137   &     129.243  \\
                              &    [0.336]   &  [0.388]    &   [ 0.022]  &    [0.507]   &    [0.393]   &  [0.480]   &    [0.115]   \\
Capital Grant&       15.362   &       0.088   &      55.631 &      75.609   &    -130.562   &       2.438   &      64.822   \\
                 &   [0.392]   &     [0.882]   &    [0.166]   &     [0.162]   &   [0.305]   &    [0.993]   &    [ 0.638]  \\
\midrule
\(N\)       &        2,320   &        2,320   &        2,320   &        2,320   &        2,320   &        2,320   &        2,320   \\
\toprule
\end{tabular}
}
\begin{minipage}{16.0cm}
\footnotesize{Standard errors in parentheses, p-values in brackets. The results in this table are based on Table 4 from Karlan et al. (2014), in year 2 Insured is instrumented using a full set of prices. Total costs (column (6)) includes sum of chemicals, land preparatory costs, hired labor, and family labor (valued at gender/community/year-specific wages). Harvest
value includes own-produced consumption, valued at community-specific market value. All specifications include controls for full set of sample frame and year interactions.
*** p \textless 0.01 ** p \textless 0.05 * p \textless 0.1.}
\end{minipage}

\end{table}

\textcolor{black}{The results of Table \ref{table:appli} are based on 2SLS regressions, while our theoretical results cover OLS ones. To alleviate this concern, we report results from reduced form regressions with only one of the instruments used by \cite{karlan2014agricultural}, the binary variable "offered a capital grant and insurance at price 0". We also include a full set of sample frame and year interactions as controls, as in their 2SLS regression. Results are shown in Table \ref{table:appli3}. Again, clustering at the region level does not change the significance of the ``intention-to-treat'' effects of that instrument.}

\begin{table}[H]
\caption{Reduced Form Effects in \cite{karlan2014agricultural}, without and with clustering. \label{table:appli3}}
	\resizebox{\textwidth}{!}{%
\begin{tabular}{l*{8}{c}}
\toprule
            &\multicolumn{1}{c}{(1)}&\multicolumn{1}{c}{(2)}&\multicolumn{1}{c}{(3)}&\multicolumn{1}{c}{(4)}&\multicolumn{1}{c}{(5)}&\multicolumn{1}{c}{(6)}&\multicolumn{1}{c}{(7)}\\
            &\multicolumn{1}{c}{ Land  }&\multicolumn{1}{c}{\# of Acres}&\multicolumn{1}{c}{  Value of }&\multicolumn{1}{c}{ Wages Paid  }&\multicolumn{1}{c}{  Opportunity Cost   }&\multicolumn{1}{c}{ Total }&\multicolumn{1}{c}{Value of }\\
            &\multicolumn{1}{c}{Preparation Costs }&\multicolumn{1}{c}{Cultivated}&\multicolumn{1}{c}{  Chemicals Used }&\multicolumn{1}{c}{ to Hired Labor }&\multicolumn{1}{c}{ of Family  Labor }&\multicolumn{1}{c}{Costs  }&\multicolumn{1}{c}{ Harvest  }\\
\midrule
\multicolumn{1}{l}{\textbf{A: Robust SE  }} \\
\addlinespace

K-Grant &  42.186** &       1.275*  &     110.629***&     115.089   &     -46.286   &     214.860   &     165.269   \\
and free insurance offered            &    (19.125)   &     (0.658)   &    (24.791)   &    (93.025)   &   (114.330)   &   (186.694)   &   (122.767)   \\
             &     [0.027]   &     [0.053]   &     [0.000]   &     [0.216]   &     [0.686]   &     [0.250]   &     [0.178]   \\

\midrule
\multicolumn{1}{l}{\textbf{B: Clustered SE }} \\
\addlinespace
K-Grant&     42.186*  &       1.275*  &     110.629***&     115.089   &     -46.286   &     214.860   &     165.269   \\
and free insurance offered            &    (20.795)   &     (0.584)   &    (23.964)   &    (95.827)   &   (110.267)   &   (192.092)   &   (150.403)   \\
            &     [0.070]   &     [0.054]   &     [0.001]   &     [0.257]   &     [0.684]   &     [0.289]   &     [0.298]   \\

\midrule
\multicolumn{1}{l}{\textbf{C: Wild Bootstrap  }} \\
\addlinespace
K-Grant &     42.186*  &       1.275***  &     110.629***&     115.089   &     -46.286   &     214.860   &     165.269   \\
  and free insurance offered           &    [0.051]   &      [0.006]  &    [0.004]  &    [0.268]   &    [0.630]   &    [0.384]   &    [0.300]   \\

\midrule
\(N\)       &        2,320   &        2,320   &        2,320   &        2,320   &        2,320   &        2,320   &        2,320   \\
\toprule
\end{tabular}
}\\
\begin{minipage}{16.0cm}
\footnotesize{Standard errors in parentheses, p-values in brackets. The results in this table are based on a reduced form regression inspired by Table 4 from Karlan et al. (2014). They consist of a regression of the outcome on a dummy variable for being offered a capital grant and insurance at price 0. Total costs (column (6)) includes sum of chemicals, land preparatory costs (e.g., equipment rental but not labor), hired labor, and family labor (valued at gender/community/year-specific wages). Harvest
value includes own-produced consumption, valued at community-specific market value. All specifications include controls for full set of sample frame and year interactions.
*** p \textless 0.01 ** p \textless 0.05 * p \textless 0.1.}
\end{minipage}

\end{table}

We compare how our clustering method performs relative to the method proposed by \cite{udrystochastic2019} to account for weather shocks. They use their method to estimate the net returns of planting-stage investments in the Ghana experiment, using the RCT treatments as instruments for investment in a 2SLS regression. Note that our results above apply to OLS regression coefficients, but we will momentarily assume they also apply to 2SLS ones, to be able to draw a comparison with the results in \cite{udrystochastic2019}. Table \ref{table:appli2} below compares three confidence intervals. The first uses the normal approximation, the estimate of returns to planting-stage investment in Table 3 Column 2 of \cite{udrystochastic2019}, and standard errors clustered at the regional level. As there are only 11 regions, the second confidence interval uses the wild-bootstrap, clustering at the region level. The third confidence interval uses the 2.5 and 97.5 percentiles from the distribution of returns to planting stage investment in Figure 5 of \cite{udrystochastic2019}. The confidence intervals clustered at the region level are much tighter than that in \cite{udrystochastic2019}. This is because those confidence intervals account for different sources of variation in the estimates. Those clustered at the regional level account for the region-level shocks that occurred over the duration of the experiment, while the confidence interval in \cite{udrystochastic2019} accounts for the variability in rainfalls over a much longer period of time.
\begin{table}[H]
\centering
 \caption{Confidence Interval for Net Returns of Planting-Stage Investments  \label{table:appli2}}
\begin{tabular}{ c c  }
\toprule

                                                  & 95\% Confidence Interval \\
Normal approximation with clustered SE at region level      &  [-64.14\%,254.79\%] \\
Wild-bootstrap with clustered SE at region level               & [-97.62\%, 289.10\%] \\
Confidence interval in \cite{udrystochastic2019}  &                                                  [-1105\% , 1509\%] \\
\toprule
\end{tabular}
\begin{minipage}{16.0cm}
\footnotesize{The results in the first two rows in this table are based on the regression in Column 2 of Table 3 of \cite{udrystochastic2019}. The confidence interval in the first row uses the normal approximation and the estimate of returns to planting-stage investment from that regression, clustering standard errors at the region level. That in the second row uses the wild cluster bootstrap and the estimate of returns to planting-stage investment from that regression, clustering standard errors at the region level. That in the third row uses the 2.5 and 97.5 percentiles from the distribution of returns to planting stage investment in Figure 5 of \cite{udrystochastic2019}.}
\end{minipage}
\end{table}

\subsection{Barriers to Household Risk Management: Evidence from India}

In this section we reexamine the results in \cite{cole2013barriers}, who conducted an experiment in India to study the effect of price and nonprice factors in the adoption of an innovative rainfall insurance product. The authors estimate the impact of the following treatments on the decision to purchase insurance: whether the household is visited by an insurance educator; whether the educator was endorsed by local agents that have close relationships with rural villages; whether the educator presented an additional  education module about the financial product; and whether the visited household received a high cash reward.\footnote{The endorsement treatment was only assigned in two-thirds of the villages.}  The treatments were assigned at the household level, within each of the 37 villages participating in their experiment. Households have time after the visit to determine whether they would like to buy insurance or not, and shocks that could affect their decision, such as weather shocks, may occur during this period.\medskip

Table \ref{table:appli1} below replicates the effects of those four treatments shown in Table 5 of \cite{cole2013barriers}.
We first use the robust variance estimators used by the authors, and then cluster standard errors at the village level. This RCT took place in 37 villages of two districts of Andhra Pradesh. Therefore, we are unable to cluster at a higher geographical level than these 37 villages, and we can only account for fairly disaggregated village-level shocks.
\begin{table}[H]
\centering
 \caption{Conditional and Unconditional Results.\label{table:appli1}}
 \resizebox{\textwidth}{!}{
\begin{tabular}{ c c c c}
\hline
     & \ \ \ Dependent Variable: Insurance Take-Up\\
\hline
                                                  & Robust s.e.      & Clustered s.e. & Cluster s.e. - Robust s.e.\\
Visit                                        & 0.115***    &    0.115  \\
                                               &  (0.043)  & (0.089) & 0.046\\

High reward                                    & 0.394*** & 0.394*** \\
                                                  & (0.034)  &  (0.045) & 0.011 \\

Education Module                                &   0.004      & 0.004 \\
                                                &  (0.032)  &  (0.037) & 0.005 \\

Endorsement                                     &    0.060       &   0.060          \\
                                                &    (0.040)              &     (0.047)    & 0.007 \\
Household controls                                    & Yes      &  Yes \\
Village FEs                                   & Yes      &   Yes \\
N                                                 & 1047     & 1047       \\
Mean of Dep Var                                   & 0.282    & 0.282     \\
\hline
\end{tabular}
}
\begin{minipage}{16.0cm}
\footnotesize{The results in this table are based on specification 3 of Table 5 from \cite{cole2013barriers}. The dependent variable in the regression is an indicator for whether the household purchased an insurance policy. The treatment variables are indicators for whether the household was visited by an insurance educator; whether the educator was endorsed by an LSA; whether the educator presented the education module; and whether the visited household received a high cash reward. Household controls are the same as in \cite{cole2013barriers}.
Robust standard errors are shown in parentheses in the first column. Standard errors clustered at the village level are shown in the second column.\\
* p\textless{}0.10 ** p\textless{}0.05  *** p\textless{}0.01.}
\end{minipage}
\end{table}

The first column of Table \ref{table:appli1} presents results using robust standard errors. The effects of the education and endorsement treatments are insignificant. The effects of the visit and high reward treatments are significant, so for both of them we can reject $ATE(\boldsymbol{\eta}(0), \boldsymbol{\eta}(1))=0$: conditional on the village-shocks that arose during the experiment, the treatment had an effect. Assumption \ref{hyp:homog-shocks} is likely to fail in this application: receiving a visit from an educator that describes features of the insurance and answers the household's questions can affect how these households respond to village-specific economic and weather shocks arising between the visit and the time when they need to make their insurance decisions. Therefore, using robust standard errors may not be appropriate to test $ATE=0$, one may instead have to use clustered standard errors.\medskip

With clustered errors, the second column of the table shows that the effect of the high-reward treatment is still significant. We can reject $ATE=0$ for that treatment: its effect does not seem to be driven by the specific village-level shocks that arose during the experiment. On the other hand, the effect of the visit treatment is no longer significant with clustering, so we cannot assert that this treatment would still have had an effect under different village-level shocks. This could be due to the fact that the effect of this treatment was driven by the specific village-shocks that occurred during the experiment. This could also be due to the fact that accounting for the village-level shocks increases the variance of the treatment effect estimator substantially, and reduces power.\medskip

The last column of the table shows that for the four treatments, the clustered standard errors are larger than the heteroskedasticity-robust ones, and the clustered standard errors are on average 1.5 times larger. There is substantial heterogeneity across outcomes: the increase in standard errors is fairly small for the education and endorsement treatments, moderate for the high reward treatment, and very large for the visit treatment.

\section{Conclusion}

In RCTs with household-level treatment assignment and household- as well as village-level shocks affecting the potential outcomes, we show that one may use heteroskedasticity-robust or village-clustered standard errors, depending on whether one wants to draw inference on the ATE conditional on the village-level shocks, or netted out of those shocks.

\newpage
\bibliographystyle{Chicago}
\bibliography{biblio}
\newpage

\section{Appendix, not for publication}

\subsection{Tables}

\begin{table}[H]
\centering
 \caption{Conditional and Unconditional Results Of Simplified Specification from \cite{cole2013barriers}.\label{table:appli4}}
\begin{tabular}{ c c c }
\hline
     & \ \ \ Dependent Variable: Insurance Take-Up\\
\hline
                                                  & Robust s.e.      & Clustered s.e.\\
Visit                                        & 0.164***    &    0.164**  \\
                                               &  (0.054)  & (0.073)\\

Household controls                                    & No      &  No \\
Village FEs                                   & Yes      &   Yes \\
N                                                 & 416     & 416       \\
\hline
\end{tabular}
\begin{minipage}{14.0cm}
\footnotesize{The results in this table is a simplified version of  the regression of Table 5 from \cite{cole2013barriers}.  The dependent variable in the regression is an indicator for whether the household purchased an insurance policy. The treatment variable is an indicator for whether the household was visited by an insurance educator and the sample is restricted to having all other treatments equal to 0. The regression includes village fixed effects.
Robust standard errors are shown in parentheses in the first column. Standard errors clustered at the village level are shown in the second column.\\
* p\textless{}0.10 ** p\textless{}0.05  *** p\textless{}0.01.}
\end{minipage}
\end{table}

\subsection{Useful Results}

Under Assumption \ref{hyp:strat_completely_randomized}, one has that for all $i,k$

\begin{equation} \label{eq:ED}
P(D_{ik}=1)=\frac{n_{1k}}{n_k},
\end{equation}
and for all $j\ne i$
\begin{equation}\label{eq:cov}
\E(D_{ik}D_{jk})=\frac{n_{1k}(n_{1k}-1)}{n_k(n_k-1)}.
\end{equation}

\subsubsection{Lemma \ref{lem:conindep}}

\begin{lem} \label{lem:conindep}
If Assumptions \ref{hyp:main_general} and \ref{hyp:mutual-indep} hold,
\begin{equation*}
 \left(\textbf{D}_k,(\epsilon_{ik}(0),\epsilon_{ik}(1))_{1\leq i \leq n_k}\right) \indep \left(\textbf{D}_{k'},(\epsilon_{ik'}(0),\epsilon_{ik'}(1))_{1\leq i \leq n_{k'}}\right) |(\boldsymbol{\eta}(0),\boldsymbol{\eta}(1))
\end{equation*}

\end{lem}

\textbf{Proof}

By Assumption \ref{hyp:mutual-indep} and Points \ref{hyp-point:treatment-indep} and \ref{hyp-point:indep-eps-eta} of Assumption \ref{hyp:main_general},
$$\left(\textbf{D}_k,(\epsilon_{ik}(0),\epsilon_{ik}(1))_{1\leq i \leq n_k}\right) \indep \left(\textbf{D}_{k'},(\epsilon_{ik'}(0),\epsilon_{ik'}(1))_{1\leq i \leq n_{k'}},\boldsymbol{\eta}(0),\boldsymbol{\eta}(1)\right).$$
Then, the result follows from the fact joint independence implies conditional independence.

\subsection{Proof of Theorem \ref{thm:cluster-conditional}}
\subsubsection{Conditional Unbiasedness of $\widehat{ATE}$}

\begin{eqnarray*}
\E\left(\widehat{ATE}_k \middle| (\boldsymbol{\eta}(0), \boldsymbol{\eta}(1)) \right)
                        &=&\E\left(\frac{1}{n_{1k}} \underset{i}{\sum}D_{ik}Y_{ik}(1) - \frac{1}{n_{0k}} \underset{i}{\sum}\left(1-D_{ik}\right)Y_{ik}(0) \middle| (\boldsymbol{\eta}(0), \boldsymbol{\eta}(1)) \right) \nonumber \\
                        &=& \frac{1}{n_{1k}}\underset{i}{\sum}\E\left(D_{ik}Y_{ik}(1) \middle| (\boldsymbol{\eta}(0), \boldsymbol{\eta}(1))\right)-\frac{1}{n_{0k}}\underset{i}{\sum}\E\left(\left(1-D_{ik}\right)Y_{ik}(0) \middle| (\boldsymbol{\eta}(0), \boldsymbol{\eta}(1))\right) \nonumber \\
                        &=&  \frac{1}{n_{1k}} \underset{i}{\sum} \E\left(D_{ik} \middle| (\boldsymbol{\eta}(0), \boldsymbol{\eta}(1))\right) \E\left(Y_{ik}(1) \middle| (\boldsymbol{\eta}(0), \boldsymbol{\eta}(1))\right) \\ & & - \frac{1}{n_{0k}} \underset{i}{\sum} \E\left(1-D_{ik} \middle| (\boldsymbol{\eta}(0), \boldsymbol{\eta}(1))\right)\E\left(Y_{ik}(0) \middle| (\boldsymbol{\eta}(0), \boldsymbol{\eta}(1))\right)  \nonumber \\
                        &=&  \frac{1}{n_{1k}} \underset{i}{\sum} \frac{n_{1k}}{n_k}(y_{ik}(1)+\eta_k(1)) - \frac{1}{n_{0k}} \underset{i}{\sum} \frac{n_{0k}}{n_k}(y_{ik}(0) +\eta_k(0)) \nonumber \\
                        &=& (\eta_k(1)-\eta_k(0)) +\frac{1}{n_k} \underset{i}{\sum} \left(y_{ik}(1)-y_{ik}(0)\right).
\end{eqnarray*}
The first equality holds because we observe $Y_i(1)$ for treated units and $Y_i(0)$ for untreated units, the third equality follows from Assumption \ref{hyp:outcomes} and Point \ref{hyp-point:treatment-indep} of Assumption \ref{hyp:main_general}, and the fourth equality follows from Assumption \ref{hyp:outcomes} and Points \ref{hyp-point:treatment-indep} and \ref{hyp-point:indep-eps-eta} of Assumption \ref{hyp:main_general} and Equation \eqref{eq:ED}.\\

Now  $\widehat{ATE} =\frac{1}{K}\underset{k=1}{\overset{K}{\sum}}\frac{n_k}{\overline{n}}\widehat{ATE}_k$, therefore :\\
\begin{eqnarray*}
\E\left(\widehat{ATE}\middle|(\boldsymbol{\eta}(0), \boldsymbol{\eta}(1))\right)&=& \frac{1}{K} \underset{k=1}{\overset{K}{\sum}} \frac{n_k}{\overline{n}} \E\left(\widehat{ATE}_k \middle| (\boldsymbol{\eta}(0), \boldsymbol{\eta}(1)) \right)\\
                                                                        &=&\frac{1}{K}\underset{k=1}{\overset{K}{\sum}}\frac{n_k}{\overline{n}}\left[(\eta_k(1)-\eta_k(0)) +\frac{1}{n_k} \underset{i}{\sum} \left(y_{ik}(1)-y_{ik}(0)\right)\right]\\
                                                                        &=&ATE(\boldsymbol{\eta}(0), \boldsymbol{\eta}(1)).
\end{eqnarray*}
\\

\subsubsection{Conditional Variance of $\widehat{ATE}$}
We begin by deriving the conditional variance of $\widehat{ATE}_k$.

We start with:
\begin{eqnarray}
V\left(\widehat{ATE}_k\middle| (\boldsymbol{\eta}(0), \boldsymbol{\eta}(1))\right)= && V\left(\E\left(\widehat{ATE}_k\middle|\textbf{D}_k,(\boldsymbol{\eta}(0), \boldsymbol{\eta}(1))\right)\middle| (\boldsymbol{\eta}(0), \boldsymbol{\eta}(1))\right)  \nonumber\\ & &+\E\left(V\left(\widehat{ATE}_k\middle|\textbf{D}_k, (\boldsymbol{\eta}(0), \boldsymbol{\eta}(1))\right)\middle| (\boldsymbol{\eta}(0), \boldsymbol{\eta}(1))\right)\label{eq:var-proof} .
\end{eqnarray}
Now begin with the first term:
\begin{eqnarray}
& &V\left(\E\left(\widehat{ATE}_k\middle|\textbf{D}_k,(\boldsymbol{\eta}(0), \boldsymbol{\eta}(1))\right) \middle| (\boldsymbol{\eta}(0), \boldsymbol{\eta}(1))\right) \nonumber \\
&=&V\left(\frac{1}{n_{1k}}\underset{i=1}{\overset{n_k}{\sum}}D_{ik}y_{ik}(1) +\eta_k(1)-\frac{1}{n_{0k}}\underset{i=1}{\overset{n_k}{\sum}}(1-D_{ik})y_{ik}(0) -\eta_k(0)\middle|(\boldsymbol{\eta}(0), \boldsymbol{\eta}(1))\right)\nonumber\\
&=&V\left(\frac{1}{n_{1k}}\underset{i=1}{\overset{n_k}{\sum}}D_{ik}y_{ik}(1) -\frac{1}{n_{0k}}\underset{i=1}{\overset{n_k}{\sum}}(1-D_{ik})y_{ik}(0) \middle|(\boldsymbol{\eta}(0), \boldsymbol{\eta}(1))\right). \label{eq:var-ate}
\end{eqnarray}
The first equality comes from the fact that:
\begin{eqnarray*}
\E\left[D_{ik}Y_{ik}(1) \middle|\textbf{D}_k,(\boldsymbol{\eta}(0), \boldsymbol{\eta}(1))\right]&=&D_{ik}\E\left[Y_{ik}(1)|\textbf{D}_k,(\boldsymbol{\eta}(0), \boldsymbol{\eta}(1))\right] \\
&=&D_{ik}\E\left[y_{ik}(1)+\epsilon_{ik}(1) +\eta_k(1) \middle|\textbf{D}_k,(\boldsymbol{\eta}(0), \boldsymbol{\eta}(1))\right] \\
&=&D_{ik}(y_{ik}(1)+\eta_k(1)).
\end{eqnarray*}

The second equality follows from Assumption \ref{hyp:outcomes}, the third equality follows from Points \ref{hyp-point:treatment-indep} and \ref{hyp-point:indep-eps-eta} of Assumption \ref{hyp:main_general}. Similarly, one can show that $\E\left((1-D_{ik})Y_{ik}(0)\middle |\textbf{D}_k,(\boldsymbol{\eta}(0), \boldsymbol{\eta}(1))\right)=(1-D_{ik})(y_{ik}(0)+ \eta_k(0))$.\\

By Point \ref{hyp-point:treatment-indep} of Assumption \ref{hyp:main_general}, $\frac{1}{n_{1k}}\underset{i=1}{\overset{n_k}{\sum}}D_{ik}y_{ik}(1) -\frac{1}{n_{0k}}\underset{i=1}{\overset{n_k}{\sum}}(1-D_{ik})y_{ik}(0)\indep(\boldsymbol{\eta}(0), \boldsymbol{\eta}(1))$, so:
\begin{eqnarray}
&&V\left(\frac{1}{n_{1k}}\underset{i=1}{\overset{n_k}{\sum}}D_{ik}y_{ik}(1) -\frac{1}{n_{0k}}\underset{i=1}{\overset{n_k}{\sum}}(1-D_{ik})y_{ik}(0) \middle| (\boldsymbol{\eta}(0), \boldsymbol{\eta}(1))\right)\nonumber\\
&=&V\left(\frac{1}{n_{1k}}\underset{i=1}{\overset{n_k}{\sum}}D_{ik}y_{ik}(1) -\frac{1}{n_{0k}}\underset{i=1}{\overset{n_k}{\sum}}(1-D_{ik})y_{ik}(0)\right).\label{eq:eq11}
\end{eqnarray}
The right hand side of the previous equation is the variance of the estimated average treatment effect in the case of deterministic potential outcomes. Then, it follows from Equations \eqref{eq:var-ate} and \eqref{eq:eq11} and from \citet{Neyman1923} that:
\begin{equation}
V\left(\E\left(\widehat{ATE}_k|\textbf{D}_k,(\boldsymbol{\eta}(0), \boldsymbol{\eta}(1))\right)\middle| (\boldsymbol{\eta}(0), \boldsymbol{\eta}(1))\right)=\frac{1}{n_{0k}} S^2_{y(0),k} + \frac{1}{n_{1k}} S^2_{y(1),k} - \frac{1}{n_k} S^2_{y(1)-y(0),k}. \label{eq:var-neyman}
\end{equation}
Moving to the second term :
\begin{eqnarray}
&\E&\left(V\left(\widehat{ATE}_k\middle|\textbf{D}_k, (\boldsymbol{\eta}(0), \boldsymbol{\eta}(1))\right)\middle| (\boldsymbol{\eta}(0), \boldsymbol{\eta}(1))\right) \nonumber\\
&=&\E\left(V\left(\frac{1}{n_{1k}}\underset{i=1}{\overset{n_k}{\sum}}D_{ik}\epsilon_{ik}(1) - \frac{1}{n_{0k}}\underset{i=1}{\overset{n_k}{\sum}}(1-D_{ik})\epsilon_{ik}(0)\middle|\textbf{D}_k, (\boldsymbol{\eta}(0), \boldsymbol{\eta}(1))\right)\middle| (\boldsymbol{\eta}(0), \boldsymbol{\eta}(1))\right) \nonumber\\
&=&\E\left(\frac{1}{n_{1k}^2}\underset{i=1}{\overset{n_k}{\sum}}D_{ik}^2\sigma^2_{1i}+\frac{1}{n_{0k}^2}\underset{i=1}{\overset{n_k}{\sum}}\left(1-D_{ik}\right)^2\sigma^2_{0i}\middle| (\boldsymbol{\eta}(0), \boldsymbol{\eta}(1))\right)
\nonumber \\
&&-E\left(\frac{1}{n_{0k}n_{1k}}Cov\left(\underset{i=1}{\overset{n_k}{\sum}}D_{ik}\epsilon_{ik}(1),\underset{i=1}{\overset{n_k}{\sum}}(1-D_{ik})\epsilon_{ik}(0)\middle|\textbf{D}_k,(\boldsymbol{\eta}(0), \boldsymbol{\eta}(1))\right)\middle|(\boldsymbol{\eta}(0), \boldsymbol{\eta}(1))\right) \nonumber\\
&=&\frac{1}{n_{1k}^2}\underset{i=1}{\overset{n_k}{\sum}}\E\left(D_{ik}^2\right)\sigma^2_{1i}+\frac{1}{n_{0k}^2}\underset{i=1}{\overset{n_k}{\sum}}\E\left[\left(1-D_{ik}\right)^2\right]\sigma^2_{0i} -\nonumber\\&&
\frac{1}{n_{0k}n_{1k}}\E\left(Cov\left(\underset{i=1}{\overset{n_k}{\sum}}D_{ik}\epsilon_{ik}(1),\underset{i=1}{\overset{n_k}{\sum}}(1-D_{ik})\epsilon_{ik}(0)\middle|\textbf{D}_k,(\boldsymbol{\eta}(0), \boldsymbol{\eta}(1))\right)\middle|(\boldsymbol{\eta}(0), \boldsymbol{\eta}(1))\right)\nonumber \\
&=&\frac{1}{n_{1k}^2}\underset{i=1}{\overset{n_k}{\sum}}\frac{n_{1k}}{n_k}\sigma^2_{1i}+\frac{1}{n_{0k}^2}\underset{i=1}{\overset{n_k}{\sum}}\frac{n_{0k}}{n_k}\sigma^2_{0i}
\nonumber\\&&-\frac{1}{n_{0k}n_{1k}}\E\left(\underset{i=1}{\overset{n_k}{\sum}}Cov\left(D_{ik}\epsilon_{ik}(1),(1-D_{ik})\epsilon_{ik}(0)\middle|\textbf{D}_k,(\boldsymbol{\eta}(0), \boldsymbol{\eta}(1))\right)\middle|(\boldsymbol{\eta}(0), \boldsymbol{\eta}(1))\right) \nonumber\\
&=& \frac{1}{n_{1k}}\overline{\sigma^2_1}_k+  \frac{1}{n_{0k}}\overline{\sigma^2_0}_k
\nonumber\\&&-\frac{1}{n_{0k}n_{1k}}\E\left(\underset{i=1}{\overset{n_k}{\sum}}D_{ik}(1-D_{ik})Cov\left(\epsilon_{ik}(1),\epsilon_{ik}(0)\middle|\textbf{D}_k,(\boldsymbol{\eta}(0), \boldsymbol{\eta}(1))\right)\middle| (\boldsymbol{\eta}(0), \boldsymbol{\eta}(1))\right) \nonumber\\
&=&\frac{1}{n_{1k}}\overline{\sigma^2_1}_k+  \frac{1}{n_{0k}}\overline{\sigma^2_0}_k. \label{eq:var-shocks}
\end{eqnarray}

The first equality holds because conditional on $\textbf{D}_k$ and $(\boldsymbol{\eta}(0), \boldsymbol{\eta}(1))$ there is no randomness in $(1-D_{ik})(y_{ik}(0)+\eta_k(0))$ and $D_{ik}(y_{ik}(1)+\eta_k(1))$. The second equality holds by Points \ref{hyp-point:finite-var-eps}, \ref{hyp-point:indep-among-eps}, \ref{hyp-point:treatment-indep}, and  \ref{hyp-point:indep-eps-eta} of Assumption \ref{hyp:main_general}, which imply that  $V\left(\epsilon_{ik}(d)|\textbf{D}_k,(\boldsymbol{\eta}(0), \boldsymbol{\eta}(1))\right)=V\left(\epsilon_{ik}(d)\right)=\sigma^2_{dik}$ and $Cov(\epsilon_{ik}(d),\epsilon_{jk}(d)|\textbf{D}_k,(\boldsymbol{\eta}(0), \boldsymbol{\eta}(1)))=Cov(\epsilon_{ik}(d),\epsilon_{jk}(d))=0$. The third equality holds by Point \ref{hyp-point:treatment-indep} of  Assumption \ref{hyp:main_general}. The fourth equality holds since $D_{ik}$ is binary  therefore $D_{ik}^2=D_{ik}$, and because  by Points \ref{hyp-point:indep-among-eps}, \ref{hyp-point:treatment-indep}, and  \ref{hyp-point:indep-eps-eta} of Assumption \ref{hyp:main_general}, $Cov(\epsilon_{ik}(1),\epsilon_{jk}(0)|\textbf{D}_k,(\boldsymbol{\eta}(0), \boldsymbol{\eta}(1)))=Cov(\epsilon_{ik}(1),\epsilon_{jk}(0))=0$. The last equality holds since $D_{ik}(1-D_{ik})=0$.\\
Combining Equations \eqref{eq:var-proof}, \eqref{eq:var-neyman}, and \eqref{eq:var-shocks} shows that:\\
\begin{equation}
V\left(\widehat{ATE}_k\middle|(\boldsymbol{\eta}(0), \boldsymbol{\eta}(1))\right)= \frac{1}{n_{0k}} S^2_{y(0),k} + \frac{1}{n_{1k}} S^2_{y(1),k} - \frac{1}{n} S^2_{y(1)-y(0),k}+\frac{1}{n_{1k}}\overline{\sigma^2_1}_k+  \frac{1}{n_{0k}}\overline{\sigma^2_0}_k.
\end{equation}

Now by Lemma \ref{lem:conindep}, for all $k\ne k'$:
\begin{equation}
Cov\left(\widehat{ATE}_k,\widehat{ATE}_{k'} \middle| (\boldsymbol{\eta}(0),\boldsymbol{\eta}(1))\right)=0. \label{eq:ATEkcov}
\end{equation}

Finally, combining the fact that $\widehat{ATE} =\frac{1}{K}\underset{k=1}{\overset{K}{\sum}}\frac{n_k}{\overline{n}}\widehat{ATE}_k$ and Equation \eqref{eq:ATEkcov}  we get:
\begin{equation*}
V\left(\widehat{ATE}\middle|(\boldsymbol{\eta}(0),\boldsymbol{\eta}(1))\right)=\frac{1}{K^2}\underset{k=1}{\overset{K}{\sum}}\left(\frac{n_k}{\overline{n}}\right)^2V\left(\widehat{ATE}_k|((\boldsymbol{\eta}(0), \boldsymbol{\eta}(1))\right).
\end{equation*}

\subsubsection{Estimating an Upper Bound for the Conditional Variance of $\widehat{ATE}$}
We start by showing that
\begin{equation}\label{eq:upper_bound_cond_var_term1}
\E\left(\frac{1}{n_{1k}} \left[\frac{1}{n_{1k}-1}\underset{i=1}{\overset{n_k}{\sum}}D_{ik}\left(Y_{ik}(1)-\overline{Y}_{1k}\right)^2\right]\middle|(\boldsymbol{\eta}(0),\boldsymbol{\eta}(1))\right)=\frac{1}{n_{1k}} S^2_{y(1),k} + \frac{1}{n_{1k}}\overline{\sigma^2_1}_k.
\end{equation}
We have:
\begin{eqnarray*}
& &\frac{1}{n_{1k}} \left[\frac{1}{n_{1k}-1}\underset{i=1}{\overset{n_k}{\sum}}D_{ik}\left(Y_{ik}(1)-\overline{Y}_{1k}\right)^2\right]\\
&=&\frac{1}{n_{1k}}\frac{1}{n_{1k}-1}\left[\underset{i=1}{\overset{n_k}{\sum}}D_{ik}Y_{ik}(1)^2-n_{1k}\overline{Y}_{1k}^2\right]\\
&=&\frac{1}{n_{1k}}\frac{1}{n_{1k}-1}\left[\underset{i=1}{\overset{n_k}{\sum}}D_{ik}\left(y_{ik}(1)^2+2\epsilon_{ik}(1)y_{ik}(1)+\epsilon^2_{ik}(1)+\eta^2_k(1)+2y_{ik}(1)\eta_k(1)+2\eta_k(1)\epsilon_{ik}(1)\right)\right]\\
& &
-\frac{1}{n_{1k}}\frac{1}{n_{1k}-1}\left[\frac{\left(\underset{i=1}{\overset{n_k}{\sum}}D_{ik}y_{ik}(1)+\underset{i=1}{\overset{n_k}{\sum}}D_{ik}\epsilon_{ik}(1)+\eta_k(1)\underset{i=1}{\overset{n_k}{\sum}}D_{ik}\right)^2}{n_{1k}}\right]\\
&=& A + B + C + D + E + F,
\end{eqnarray*}
with
\begin{eqnarray*}
A&=&\frac{1}{n_{1k}}\frac{1}{n_{1k}-1}\left[\underset{i=1}{\overset{n_k}{\sum}}D_{ik}y_{ik}(1)^2-\frac{\left(\underset{i=1}{\overset{n_k}{\sum}}D_{ik}y_{ik}(1)\right)^2}{n_{1k}}\right],\\
B&=&\frac{1}{n_{1k}}\frac{1}{n_{1k}-1}\left[\underset{i=1}{\overset{n_k}{\sum}}D_{ik}\epsilon^2_{ik}(1)-\frac{\left(\underset{i=1}{\overset{n_k}{\sum}}D_{ik}\epsilon_{ik}(1)\right)^2}{n_{1k}}\right],\\
C&=&\frac{1}{n_{1k}}\frac{2}{n_{1k}-1}\left[\underset{i=1}{\overset{n_k}{\sum}}D_{ik}y_{ik}(1)\epsilon_{ik}(1)-\frac{\left(\left(\underset{i=1}{\overset{n_k}{\sum}}D_{ik}y_{ik}(1)\right)\left(\underset{i=1}{\overset{n_k}{\sum}}D_{ik}\epsilon_{ik}(1)\right)\right)}{n_{1k}}\right],
\end{eqnarray*}

and

\begin{eqnarray*}
D&=&\frac{1}{n_{1k}}\frac{1}{n_{1k}-1}\left[\eta^2_k(1)\underset{i=1}{\overset{n_k}{\sum}}D_{ik}-\frac{1}{n_{1k}}\left(\eta_k(1)\underset{i=1}{\overset{n_k}{\sum}}D_{ik}\right)^2\right]=0,\\
E&=&\frac{1}{n_{1k}}\frac{1}{n_{1k}-1}\left[2\eta_k(1)\underset{i=1}{\overset{n_k}{\sum}}D_{ik}y_{ik}(1)-\frac{2}{n_{1k}}\left(\underset{i=1}{\overset{n_k}{\sum}}D_{ik}y_{ik}(1)\right)\left(\eta_k(1)\underset{i=1}{\overset{n_k}{\sum}}D_{ik}\right)\right]=0,\\
F&=&\frac{1}{n_{1k}}\frac{1}{n_{1k}-1}\left[2\eta_k(1)\underset{i=1}{\overset{n_k}{\sum}}D_{ik}\epsilon_{ik}(1)-\frac{2}{n_{1k}}\left(\underset{i=1}{\overset{n_k}{\sum}}D_{ik}\epsilon_{ik}(1)\right)\left(\eta_k(1)\underset{i=1}{\overset{n_k}{\sum}}D_{ik}\right)\right]=0.
\end{eqnarray*}
To ease notation let $\E^*(X)=\E(X |(\boldsymbol{\eta}(0),\boldsymbol{\eta}(1)))$.
\begin{eqnarray*}
\E^*(A)&=&\E(A)\\
&=&\frac{1}{n_{1k}(n_{1k}-1)}\left[\underset{i=1}{\overset{n_k}{\sum}}\E(D_{ik})y_{ik}^2(1)-\frac{1}{n_{1k}}\underset{i=1}{\overset{n_k}{\sum}}\E(D_{ik})y_{ik}^2(1)-\frac{1}{n_{1k}}\underset{i\neq j}{\sum\sum}\E(D_{ik}D_{jk})y_{ik}(1)y_{jk}(1)\right]\\
&=&\frac{1}{n_{1k}(n_{1k}-1)}\left[\frac{n_{1k}}{n_k}\underset{i=1}{\overset{n_k}{\sum}}y_{ik}^2(1)-\frac{1}{n_{1k}}\underset{i=1}{\overset{n_k}{\sum}}\frac{n_{1k}}{n_k}y_{ik}^2(1)-\frac{1}{n_{1k}}\underset{i\neq j}{\sum\sum}\frac{n_{1k}}{n_k}\frac{n_{1k}-1}{n_k-1}y_{ik}(1)y_{jk}(1)\right]\\
&=&\frac{1}{n_{1k}(n_{1k}-1)}\left[\frac{n_{1k}-1}{n_k} \underset{i=1}{\overset{n_k}{\sum}}y_{ik}^2(1) - \frac{n_{1k}-1}{n_k(n_k-1)}\underset{i\neq j}{\sum\sum}y_{ik}(1)y_{jk}(1)\right]\\
&=&\frac{1}{n_{1k}(n_{1k}-1)}\left[\frac{n_{1k}-1}{n_k} \underset{i=1}{\overset{n_k}{\sum}}y_{ik}^2(1)+\frac{n_{1k}-1}{n_k(n_k-1)} \underset{i=1}{\overset{n_k}{\sum}}y_{ik}^2(1) - \frac{n_{1k}-1}{n_k(n_k-1)}\underset{i=1}{\overset{n_k}{\sum}}\underset{j=1}{\overset{n_k}{\sum}}y_{ik}(1)y_{jk}(1)\right]\\
&=&\frac{1}{n_{1k}(n_{1k}-1)}\left[\frac{(n_{1k}-1)(n_k-1)+(n_{1k}-1)}{n_k(n_k-1)}\underset{i=1}{\overset{n_k}{\sum}}y_{ik}^2(1)-\frac{n_k(n_{1k}-1)}{(n_k-1)}\underset{i=1}{\overset{n_k}{\sum}}\underset{j=1}{\overset{n_k}{\sum}}\frac{y_{ik}(1)}{n_k}\frac{y_{jk}(1)}{n_k}\right]\\
&=&\frac{1}{n_{1k}}\left[\frac{1}{n_k-1}\underset{i=1}{\overset{n_k}{\sum}}y_{ik}^2(1)-\frac{n_k}{n_k-1}\overline{y}_k^2(1)\right]\\
&=& \frac{1}{n_{1k}} S^2_{y(1),k}.
\end{eqnarray*}
The first equality holds by Point \ref{hyp-point:treatment-indep} of Assumption \ref{hyp:main_general}.
The third holds by Equations \eqref{eq:ED} and \eqref{eq:cov}.

Moving to $B$,
\begin{eqnarray*}
\E^*(B)&=&\frac{1}{n_{1k}(n_{1k}-1)}\left[\underset{i=1}{\overset{n_k}{\sum}}\E^*(D_{ik})\E^*(\epsilon^2_{ik}(1))-\frac{\E^*\left(\left(\underset{i=1}{\overset{n_k}{\sum}}D_{ik}\epsilon_{ik}(1)\right)^2\right)}{n_{1k}}\right]\\
&=&\frac{1}{n_{1k}(n_{1k}-1)}\Bigg[\underset{i=1}{\overset{n_k}{\sum}}\E^*(D_{ik})\E^*(\epsilon^2_{ik}(1))-\frac{1}{n_{1k}}\underset{i=1}{\overset{n_k}{\sum}}\E^*(D_{ik}^2)\E^*(\epsilon_{ik}^2(1))\\
  &&-\frac{1}{n_{1k}}\underset{i\neq j}{\sum\sum}\E^*(D_{ik}D_{jk})\E^*(\epsilon_{ik}(1)\epsilon_{jk}(1))\Bigg]\\
&=&\frac{1}{n_{1k}(n_{1k}-1)}\left[\frac{n_{1k}}{n_k}\underset{i=1}{\overset{n_k}{\sum}}\sigma_{1i}^2-\frac{1}{n_{1k}}\underset{i=1}{\overset{n_k}{\sum}}\frac{n_{1k}}{n_k}\sigma_{1i}^2\right]\\
&=&\frac{1}{n_{1k}}\overline{\sigma^2_1}_k.
\end{eqnarray*}

The first equality holds by Points \ref{hyp-point:finite-var-eps} and \ref{hyp-point:treatment-indep} of Assumption \ref{hyp:main_general}. The second equality holds by  Points  \ref{hyp-point:finite-var-eps} and \ref{hyp-point:treatment-indep} of Assumption \ref{hyp:main_general}.
The third equality holds since $D_{ik}^2=D_{ik}$ and by Points \ref{hyp-point:finite-var-eps}, \ref{hyp-point:indep-among-eps}, \ref{hyp-point:treatment-indep}, and \ref{hyp-point:indep-eps-eta} of Assumption \ref{hyp:main_general} as well as  by Equation \eqref{eq:ED}, which imply $\E^*(D_{ik})=\E(D_{ik})=\frac{n_{1k}}{n_k}$, $E^*(\epsilon^2_{ik}(1))=E(\epsilon^2_{ik}(1))=\sigma_{1i}^2$, and $\E^*(\epsilon_{ik}(1)\epsilon_{jk}(1))=\E(\epsilon_{ik}(1)\epsilon_{jk}(1))=0$.\\

Finally for $C$:
\begin{eqnarray*}
\E^*(C)&=&\frac{1}{n_{1k}}\frac{2}{n_{1k}-1}\left[\underset{i=1}{\overset{n_k}{\sum}}\E^*(D_{ik})y_{ik}(1)\E^*(\epsilon_{ik}(1))-\frac{\E^*\left(\underset{i=1}{\overset{n_k}{\sum}}D_{ik}y_{ik}(1)\underset{i=1}{\overset{n_k}{\sum}}D_{ik}\epsilon_{ik}(1)\right)}{n_{1k}}\right]\\
&=& \frac{1}{n_{1k}}\frac{2}{n_{1k}-1}\left[-\underset{i=1}{\overset{n_k}{\sum}}\E^*(D_{ik}^2)y_{ik}(1)\E^*(\epsilon_{ik}(1))-\underset{i\neq j}{\sum\sum}\E^*(D_{ik}D_{jk})y_{ik}(1)\E^*(\epsilon_{jk}(1))\right]\\
&=&0.
\end{eqnarray*}

The first  equality holds by Point \ref{hyp-point:treatment-indep} of Assumption \ref{hyp:main_general}. The second and third equalities hold by Point \ref{hyp-point:indep-eps-eta} of Assumption \ref{hyp:main_general}, which implies $\E^*(\epsilon_{ik}(1))=\E(\epsilon_{ik}(1))=0$. This completes the proof of \eqref{eq:upper_bound_cond_var_term1}.
Using similar arguments, one can show that
\begin{equation}\label{eq:upper_bound_cond_var_term2}
\E\left(\frac{1}{n_{0k}} \left[\frac{1}{n_{0k}-1}\underset{i=1}{\overset{n_k}{\sum}}\left(1-D_{ik}\right)\left(Y_{ik}(0)-\overline{Y}_{0k}\right)^2\right]\middle|(\boldsymbol{\eta}(0),\boldsymbol{\eta}(1))\right)=\frac{1}{n_{0k}} S^2_{y(0),k} + \frac{1}{n_{0k}}\overline{\sigma^2_0}_k.
\end{equation}
Point 2 of the theorem and Equations \eqref{eq:upper_bound_cond_var_term1} and \eqref{eq:upper_bound_cond_var_term2} imply that
$$V\left(\widehat{ATE}_k|(\boldsymbol{\eta}(0),\boldsymbol{\eta}(1))\right)\leq  \E\left(\widehat{V}_{rob}\left(\widehat{ATE}_k\right)\middle|(\boldsymbol{\eta}(0),\boldsymbol{\eta}(1))\right).$$
Finally, the result follows from Point 2 of the theorem, the definition of $\widehat{V}_{rob}\left(\widehat{ATE}\right)$, and the linearity of the conditional expectation.

\subsection{Proof of Theorem  \ref{thm:main-cluster}}
\subsubsection{Unbiasedness of $\widehat{ATE}$}
\begin{eqnarray}
\E\left(\widehat{ATE}\right)&=& \E\left[\E\left(\widehat{ATE}|\boldsymbol{\eta}(0), \boldsymbol{\eta}(1)\right)\right] \nonumber \\
                        &=& \E\left[ATE(\boldsymbol{\eta}(0), \boldsymbol{\eta}(1)) \right]\nonumber \\
                        &=& \E\left[\frac{1}{K} \underset{k=1}{\overset{K}{\sum}}\frac{n_k}{\overline{n}} \underset{i=0}{\overset{n_k}{\sum}}\left[(y_{ik}(1)+\eta_k(1))-(y_{ik}(0)+\eta_k(0))\right]\right] \nonumber\\
                        &=& \frac{1}{K} \underset{k=1}{\overset{K}{\sum}}\frac{n_k}{\overline{n}} \underset{i=0}{\overset{n_k}{\sum}}\left[y_{ik}(1)-y_{ik}(0)\right] \nonumber \\
                        &=& ATE.
\end{eqnarray}

Where the first equality holds by the law of iterated expectations, and the second equality holds by Theorem \ref{thm:cluster-conditional}.

\subsubsection{Unconditional Variance of $\widehat{ATE}$}
We start with:
\begin{equation}
V\left(\widehat{ATE}\right)=\E\left(V\left(\widehat{ATE}|\boldsymbol{\eta}(0), \boldsymbol{\eta}(1)\right)\right)+ V\left(\E\left(\widehat{ATE}|\boldsymbol{\eta}(0), \boldsymbol{\eta}(1)\right)\right). \label{eq:3forcor}
\end{equation}
Now begin with the first term:
\begin{eqnarray}
\E\left(V\left(\widehat{ATE}|\boldsymbol{\eta}(0), \boldsymbol{\eta}(1)\right)\right)&=&\E\left[\frac{1}{K^2}\underset{k=1}{\overset{K}{\sum}}\left(\frac{n_k}{\overline{n}}\right)^2V\left(\widehat{ATE}_k|\boldsymbol{\eta}(0), \boldsymbol{\eta}(1)\right)\right] \nonumber\\
&=&\frac{1}{K^2}\underset{k=1}{\overset{K}{\sum}}\left(\frac{n_k}{\overline{n}}\right)^2V\left(\widehat{ATE}_k|\boldsymbol{\eta}(0), \boldsymbol{\eta}(1)\right). \label{eq:0forcor}
\end{eqnarray}

Where the first equality holds by Theorem \ref{thm:cluster-conditional}, and the second equality holds because \\ $V\left(\widehat{ATE}_k|\boldsymbol{\eta}(0), \boldsymbol{\eta}(1)\right)$ contains no stochastic components, as shown in Point 2 of Theorem \ref{thm:cluster-conditional}.\\

Moving to the second term:
\begin{eqnarray}
& &V\left(\E\left(\widehat{ATE}|\boldsymbol{\eta}(0), \boldsymbol{\eta}(1)\right)\right) \nonumber\\
&=&V\left(ATE(\boldsymbol{\eta}(0), \boldsymbol{\eta}(1)) \right)  \nonumber\\
&=&V\left(\frac{1}{K} \underset{k=1}{\overset{K}{\sum}}\frac{n_k}{\overline{n}} \frac{1}{n_k} \underset{i=0}{\overset{n_k}{\sum}}\left[(y_{ik}(1)+\eta_k(1))-(y_{ik}(0)+\eta_k(0))\right]\right) \nonumber\\
&=&\frac{1}{K^2}V\left(\underset{k=1}{\overset{K}{\sum}}\frac{1}{\overline{n}}\underset{i=0}{\overset{n_k}{\sum}}(y_{ik}(1)-y_{ik}(0))+(\eta_k(1)-\eta_k(0))\right) \nonumber\\
&=&\frac{1}{K^2}V\left(\underset{k=1}{\overset{K}{\sum}}\frac{n_k}{\overline{n}}(\eta_k(1)-\eta_k(0))\right) \nonumber\\
&=&\frac{1}{K^2}\underset{k=1}{\overset{K}{\sum}}\left(\frac{n_k}{\overline{n}}\right)^2V\left(\eta_k(1)-\eta_k(0)\right). \label{eq:2forcor}
\end{eqnarray}
The first equality holds by Theorem \ref{thm:cluster-conditional}, the last holds by Assumption \ref{hyp:mutual-indep} and Point \ref{hyp-point:finite-var-etas} of Assumption \ref{hyp:main_general}. The result follows from \eqref{eq:3forcor}, \eqref{eq:0forcor}, and \eqref{eq:2forcor}, and from Point 2 of Theorem \ref{thm:cluster-conditional}.

\subsubsection{Estimating an Upper Bound for the Variance of $\widehat{ATE}$ \label{assum1uncon}}

The first inequality is trivial so we only prove the second one.
\begin{eqnarray}
\E\left[\widehat{V}_{clu}(\widehat{ATE})\right]&= &\E\left[\frac{1}{K\left(K-1\right)}\underset{k=1}{\overset{K}{\sum}}\left(\frac{n_k}{\overline{n}}\widehat{ATE}_k - \widehat{ATE}\right)^2\right] \nonumber\\
&=&\frac{1}{K\left(K-1\right)}\E\left[\underset{k=1}{\overset{K}{\sum}}\left(\frac{n_k}{\overline{n}}\right)^2\widehat{ATE}_k^2-2\widehat{ATE}\underset{k=1}{\overset{K}{\sum}}\frac{n_k}{\overline{n}}\widehat{ATE}_k + K\widehat{ATE}^2\right] \nonumber \\
&=&\frac{1}{K\left(K-1\right)}\left[\underset{k=1}{\overset{K}{\sum}}\left(\frac{n_k}{\overline{n}}\right)^2\E\left(\widehat{ATE}_k^2\right)-K\E\left(\widehat{ATE}^2\right)\right] \nonumber \\
&=&\frac{1}{K\left(K-1\right)}\left[\underset{k=1}{\overset{K}{\sum}}\left(\frac{n_k}{\overline{n}}\right)^2\left[V\left(\widehat{ATE}_k\right)+\E\left(\widehat{ATE}_k\right)^2\right]-K\left[V\left(\widehat{ATE}\right)+\E\left(\widehat{ATE}\right)^2\right]\right] \nonumber\\
&=&\frac{1}{K\left(K-1\right)}\left[K^2V\left(\widehat{ATE}\right)+\underset{k=1}{\overset{K}{\sum}}\left(\frac{n_k}{\overline{n}}\right)^2\E\left(\widehat{ATE}_k\right)^2-KV\left(\widehat{ATE}\right)-K\E\left(\widehat{ATE}\right)^2\right] \nonumber\\
&=&V\left(\widehat{ATE}\right)+\frac{1}{K\left(K-1\right)}\left[\underset{k=1}{\overset{K}{\sum}}\left(\frac{n_k}{\overline{n}}\right)^2\E\left(\widehat{ATE}_k\right)^2-K\E\left(\widehat{ATE}\right)^2\right]. \label{eq:1forcor}
\end{eqnarray}
The second and third equalities follow from algebraic manipulations and the linearity of the expectations operator, the fourth follows from the definition of a variance, and the fifth follows from $V\left(\widehat{ATE}\right)=\frac{1}{K^2}\underset{k=1}{\overset{K}{\sum}}\left(\frac{n_k}{\overline{n}}\right)^2V\left(\widehat{ATE}_k\right)$.\\

By convexity of $x \rightarrow x^2$,
\begin{eqnarray*}
&\frac{1}{K}\underset{k=1}{\overset{K}{\sum}}\left(\frac{n_k}{\overline{n}}\right)^2\E\left(\widehat{ATE}_k\right)^2\geq\left[\frac{1}{K}\underset{k=1}{\overset{K}{\sum}}\left(\frac{n_k}{\overline{n}}\right)\E\left(\widehat{ATE}_k\right)\right]^2\\
\Leftrightarrow&\underset{k=1}{\overset{K}{\sum}}\left(\frac{n_k}{\overline{n}}\right)^2\E\left(\widehat{ATE}_k\right)^2\geq K\E\left(\widehat{ATE}\right)^2,
\end{eqnarray*}
so the second term in Equation \eqref{eq:1forcor} is positive. This proves the result.

\subsection{Proof of Corollary  \ref{cor:vars}}

If $n_k=\overline{n}$, it follows from Equation \eqref{eq:1forcor} that
\begin{equation*}
\E\left[\widehat{V}_{clu}(\widehat{ATE})\right] = V\left(\widehat{ATE}\right)+\frac{1}{K\left(K-1\right)}\left[\underset{k=1}{\overset{K}{\sum}}E\left(\widehat{ATE}_k\right)^2-K\E\left(\widehat{ATE}\right)^2\right].
\end{equation*}
Moreover,
\begin{align*}
&\E\left[\widehat{V}_{rob}\left(\widehat{ATE}_k\right)\right]\\
=&\E\left[\E^*\left(\widehat{V}_{rob}\left(\widehat{ATE}_k\right)\right)\right]\\
=&\E\left[\E^*\left(\frac{1}{K^2}\underset{k=1}{\overset{K}{\sum}}\left(\frac{n_k}{\overline{n}}\right)\widehat{V}_{rob}\left(\widehat{ATE}_k\right)\right)\right]\\
=&\E\left[V\left(\widehat{ATE}|(\boldsymbol{\eta}(0),\boldsymbol{\eta}(1))\right) + \frac{1}{K^2}\underset{k=1}{\overset{K}{\sum}}  \frac{1}{n_k}\left(\frac{n_k}{\overline{n}}\right)^2 S^2_{y(1)-y(0),k} \right]\\
=&\E\left[V\left(\widehat{ATE}|(\boldsymbol{\eta}(0),\boldsymbol{\eta}(1))\right)\right] + \frac{1}{K^2}\underset{k=1}{\overset{K}{\sum}} \frac{1}{n_k} S^2_{y(1)-y(0),k},
\end{align*}
where the first equality holds by the law of iterated expectations, the third equality holds by Equations \eqref{eq:upper_bound_cond_var_term1} and \eqref{eq:upper_bound_cond_var_term2} and Point 2 of Theorem \ref{thm:cluster-conditional}, and the last follows from $n_k=\overline{n}$. Combining the two preceding displays,
\begin{align*}
&\E\left[\widehat{V}_{clu}(\widehat{ATE})\right]-\E\left[\widehat{V}_{rob}\left(\widehat{ATE}_k\right)\right]\\
=& V\left(\widehat{ATE}\right)-\E\left[V\left(\widehat{ATE}|(\boldsymbol{\eta}(0),\boldsymbol{\eta}(1))\right)\right]+ \\ &\frac{1}{K\left(K-1\right)}\left[\underset{k=1}{\overset{K}{\sum}}E\left(\widehat{ATE}_k\right)^2-K\E\left(\widehat{ATE}\right)^2\right]- \frac{1}{K^2}\underset{k=1}{\overset{K}{\sum}} \frac{1}{n_k} S^2_{y(1)-y(0),k} \\
=&\frac{1}{K^2}\underset{k=1}{\overset{K}{\sum}}V\left(\eta_k(1)-\eta_k(0)\right)+\frac{1}{K\left(K-1\right)}\left[\underset{k=1}{\overset{K}{\sum}}E\left(\widehat{ATE}_k\right)^2-K\E\left(\widehat{ATE}\right)^2\right]- \frac{1}{K^2}\underset{k=1}{\overset{K}{\sum}} \frac{1}{n_k} S^2_{y(1)-y(0),k},
\end{align*}
where the second equality follows from Equations \eqref{eq:3forcor} and \eqref{eq:2forcor} and $n_k=\overline{n}$. This proves the result.

\subsection{Proof of Theorem  \ref{thm:asym-uncon}}

Let:
\begin{eqnarray*}
&&\frac{1}{S_K/K}\left[\widehat{ATE}-ATE\right]\\
&=&\frac{1}{S_K}K\left[\widehat{ATE}-\E\left(\widehat{ATE}\right)\right]\\
&=& \frac{1}{S_K}K\left[\frac{1}{K}\sum_{k=1}^{K}AD_k-\E\left(\frac{1}{K}\sum_{k=1}^{K}AD_k\right)\right]\\
&=&\frac{1}{S_K}\sum_{k=1}^{K}\left[AD_k-\E\left(AD_k\right)\right],\\
\end{eqnarray*}

where the first equality holds by Theorem \ref{thm:main-cluster}. Under Assumptions \ref{hyp:outcomes}, \ref{hyp:main_general}, \ref{hyp:strat_completely_randomized}, \ref{hyp:mutual-indep}, and  \ref{hyp:reg-cond}, $\sum_{k=1}^{K}\left[AD_k-\E\left(AD_k\right)\right]$ is a sum of independent mean 0 random variables with finite variance. Furthermore, by Point 3 of Assumption \ref{hyp:reg-cond} we know that $\underset{K\rightarrow+\infty}{\lim}\frac{1}{S_K^{2+\epsilon}}\sum_{k=1}^{K}\E\left[|AD_k-E(AD_k)|^{2+\epsilon}\right]=0$ for some $\epsilon>0$. Therefore by the Lyapunov CLT:
\begin{equation*}
\frac{1}{S_K}\sum_{k=1}^{K}\left[AD_k-\E\left(AD_k\right)\right] \overset{d}{\rightarrow} N(0,1).
\end{equation*}

Now note that:
\begin{eqnarray*}
\frac{1}{S_K/K}\left[\widehat{ATE}-ATE\right]&=&\frac{\sqrt{K}}{\sqrt{\frac{1}{K}\sum_{k=1}^{K}V\left(AD_k\right)}}\left[\widehat{ATE}-ATE\right]\\
&=&\frac{\sqrt{K}}{\sqrt{KV\left(\widehat{ATE}\right)}}\left[\widehat{ATE}-ATE\right],
\end{eqnarray*}
so by the Slutsky Lemma and Point 3 of Assumption \ref{hyp:reg-cond}:
\begin{equation*}
\sqrt{K}\left(\widehat{ATE}-ATE\right)\overset{d}{\rightarrow} N\left(0,\sigma^2\right).
\end{equation*}
Now we show that  $K\widehat{V}_{clu}\left(\widehat{ATE}\right) \overset{p}{\rightarrow} \sigma_+^2 \geq \sigma^2$:
\begin{eqnarray*}
&& \lim_{K\rightarrow \infty} K\widehat{V}_{clu}\left(\widehat{ATE}\right)\\
&=&\lim_{K\rightarrow \infty} \frac{1}{K}\sum_{k=1}^{K}\left(AD_k-\widehat{ATE}\right)^2\\
&=&\lim_{K\rightarrow \infty} \frac{1}{K}\sum_{k=1}^{K}AD_k^2 -\left(\frac{1}{K}\sum_{k=1}^K AD_k\right)^2.
\end{eqnarray*}
By  Assumption \ref{hyp:mutual-indep}, Point 1 of Assumption \ref{hyp:reg-cond}, the strong law of large numbers in Lemma 1 of \cite{liu1988bootstrap}, the fact that almost sure convergence implies convergence in probability, and Point 3 of Assumption \ref{hyp:reg-cond},
\begin{eqnarray*}
&&\frac{1}{K}\sum_{k=1}^{K}AD_k^2 \overset{p}{\rightarrow} \lim_{K\rightarrow\infty} \frac{1}{K}\sum_{k=1}^{K}\E\left(AD_k^2\right)   \\
&&\frac{1}{K}\sum_{k=1}^{K}AD_k \overset{p}{\rightarrow}  \lim_{K\rightarrow\infty} \frac{1}{K}\sum_{k=1}^{K}\E\left(AD_k\right).
\end{eqnarray*}
Therefore, by the continuous mapping theorem:
\begin{equation*}
K\widehat{V}_{clu}\left(\widehat{ATE}\right)  \overset{p}{\rightarrow}\lim_{K\rightarrow \infty}  \frac{1}{K}\sum_{k=1}^{K}\E\left(AD_k^2\right) - \left(\frac{1}{K}\sum_{k=1}^{K}\E\left(AD_k\right)\right)^2=\sigma_+^2    .
\end{equation*}
By the convexity of $x\rightarrow x^2$ we have $\frac{1}{K}\sum_{k=1}^{K}\E\left(AD_k\right)^2 \geq \left(\frac{1}{K}\sum_{k=1}^{K}\E\left(AD_k\right)\right)^2$, so $\sigma_+^2 \geq \sigma^2$.

\subsection{Relaxing Assumption \ref{hyp:outcomes} \label{section-noassum1}}

Let $Y_{ik}(d)=f_{ikd}\left(\epsilon_{ik}(d),\eta_k(d)\right)$ for some functions $f_{ikd}(.)$. Redefine $ATE(\boldsymbol{\eta}(0), \boldsymbol{\eta}(1))$  as $\frac{1}{n}\sum_{i,k}E(Y_{ik}(1)-Y_{ik}(0)|(\boldsymbol{\eta}(0), \boldsymbol{\eta}(1)))$, and $ATE$ as $\frac{1}{n}\sum_{i,k}E(Y_{ik}(1)-Y_{ik}(0))$. It is trivial to show that $\widehat{ATE}$ is unbiased for $ATE$ and conditionally unbiased for $E(Y_{ik}(1)-Y_{ik}(0)|(\boldsymbol{\eta}(0), \boldsymbol{\eta}(1)))$, so we will focus on showing that our results regarding the variance and conditional variance of  $\widehat{ATE}$ still hold.

\subsubsection{Conditional Variance of $\widehat{ATE}$ \label{noassum1con}}

\eqref{eq:var-proof} still holds. Starting with the first term in \eqref{eq:var-proof}:
\begin{align*}
&V\left(\E\left(\widehat{ATE}_k \middle| \textbf{D}_k ,(\boldsymbol{\eta}(0), \boldsymbol{\eta}(1))\right) \middle| (\boldsymbol{\eta}(0), \boldsymbol{\eta}(1))\right)\\
&=V\left(\frac{1}{n_{1k}}\sum_{i=1}^{n_{1k}}D_{ik}\E(Y_{ik}(1)|(\boldsymbol{\eta}(0),\boldsymbol{\eta}(1)))-\frac{1}{n_{0k}}\sum_{i=1}^{n_{0k}}(1-D_{ik})\E(Y_{ik}(0)|(\boldsymbol{\eta}(0), \boldsymbol{\eta}(1)))\middle|(\boldsymbol{\eta}(0),\boldsymbol{\eta}(1))\right)
\end{align*}
where the equality holds by Point \ref{hyp-point:treatment-indep} of Assumption \ref{hyp:main_general}. Now note that conditional on $(\boldsymbol{\eta}(0), \boldsymbol{\eta}(1))$, only the $D_{ik}$s are random inside the conditional variance operator. Then, it follows from \citet{Neyman1923} that
\begin{equation*}
V\left(\E\left(\widehat{ATE}_k|\textbf{D}_k,(\boldsymbol{\eta}(0), \boldsymbol{\eta}(1))\right)\middle| (\boldsymbol{\eta}(0), \boldsymbol{\eta}(1))\right)=\frac{1}{n_{0k}} S^2_{\E^*(Y_{ik}(0)),k} + \frac{1}{n_{1k}} S^2_{\E^*(Y_{ik}(1)),k} - \frac{1}{n_k} S^2_{\E^*(Y_{ik}(1))-\E^*(Y_{ik}(0)),k}
\end{equation*}
where as before $\E^*(X)=\E(X |(\boldsymbol{\eta}(0),\boldsymbol{\eta}(1)))$. Moving to the second term in \eqref{eq:var-proof},
\begin{align*}
&\E\left(V\left(\widehat{ATE}_k\middle| \textbf{D}_k,(\boldsymbol{\eta}(0), \boldsymbol{\eta}(1))\right) \middle| (\boldsymbol{\eta}(0), \boldsymbol{\eta}(1))\right)\\
&= \E\left(\frac{1}{n_{1k}^2}V\left(\sum_{i=1}^{n_{1k}}D_{ik}Y_{ik}(1)\middle|\textbf{D}_k,(\boldsymbol{\eta}(0),\boldsymbol{\eta}(1))\right)+ \frac{1}{n_{0k}^2}V\left(\sum_{i=1}^{n_{1k}}(1-D_{ik})Y_{ik}(0)\middle|\textbf{D}_k,(\boldsymbol{\eta}(0),\boldsymbol{\eta}(1))\right)  \middle| (\boldsymbol{\eta}(0), \boldsymbol{\eta}(1))\right) \\
&=\E\left(\frac{1}{n_{1k}^2}\sum_{i=1}^{n_{1k}}D_{ik}V\left(Y_{ik}(1)\middle|\boldsymbol{\eta}(0),\boldsymbol{\eta}(1))\right)+ \frac{1}{n_{0k}^2}\sum_{i=1}^{n_{0k}}(1-D_{ik})V\left(Y_{ik}(0)\middle|\boldsymbol{\eta}(0),\boldsymbol{\eta}(1)\right)  \middle| (\boldsymbol{\eta}(0), \boldsymbol{\eta}(1))\right) \\
&=\frac{1}{n_{1k}^2}\sum_{i=1}^{n_{1k}}\E^*\left(D_{ik}\right)V\left(Y_{ik}(1)\middle|(\boldsymbol{\eta}(0),\boldsymbol{\eta}(1))\right)+ \frac{1}{n_{0k}^2}\sum_{i=1}^{n_{0k}}\E^*\left(1-D_{ik}\right)V\left(Y_{ik}(0)\middle|(\boldsymbol{\eta}(0),\boldsymbol{\eta}(1))\right) \\
&=\frac{1}{n_{1k}}\sum_{i=1}^{n_{1k}}\frac{1}{n_k}V\left(Y_{ik}(1)\middle|(\boldsymbol{\eta}(0),\boldsymbol{\eta}(1))\right)+ \frac{1}{n_{0k}}\sum_{i=1}^{n_{0k}}\frac{1}{n_k}V\left(Y_{ik}(0)\middle|(\boldsymbol{\eta}(0),\boldsymbol{\eta}(1))\right)
\end{align*}

where the second equality holds because $V\left(Y_{ik}(d)\middle|\textbf{D}_k,\boldsymbol{\eta}(0),\boldsymbol{\eta}(1)\right)=V\left(Y_{ik}(d)\middle|\boldsymbol{\eta}(0),\boldsymbol{\eta}(1)\right)$ by Point \ref{hyp-point:treatment-indep} of Assumption \ref{hyp:main_general}, and $cov\left(Y_{ik}(d),Y_{jk}(d)\middle|\textbf{D}_k,\boldsymbol{\eta}(0),\boldsymbol{\eta}(1)\right)=cov\left(Y_{ik}(d),Y_{jk}(d)\middle|\boldsymbol{\eta}(0),\boldsymbol{\eta}(1)\right)=0$, by Point \ref{hyp-point:treatment-indep} of Assumption \ref{hyp:main_general} and because $\epsilon_{ik}(d)\indep \epsilon_{jk}(d)|\boldsymbol{\eta}(0),\boldsymbol{\eta}(1)$ by Points \ref{hyp-point:indep-among-eps} and \ref{hyp-point:indep-eps-eta} of Assumption \ref{hyp:main_general}. The third equality holds because $V\left(Y_{ik}(1)\middle|(\boldsymbol{\eta}(0),\boldsymbol{\eta}(1))\right)$ and $V\left(Y_{ik}(0)\middle|(\boldsymbol{\eta}(0),\boldsymbol{\eta}(1))\right)$ are functions of $(\boldsymbol{\eta}(0),\boldsymbol{\eta}(1))$. The fourth equality holds because $\E^*\left(D_{ik}\right)=\E\left(D_{ik}\right)$ by Point \ref{hyp-point:treatment-indep} of Assumption  \ref{hyp:main_general}. Finally, the first equality holds because:
\begin{align*}
&\frac{1}{n_{1k}n_{0k}}Cov\left(\sum_{i=1}^{n_{k}}D_{ik}Y_{ik}(1),\sum_{i=1}^{n_{k}}(1-D_{ik})Y_{ik}(0)\middle|\textbf{D}_k,(\boldsymbol{\eta}(0),\boldsymbol{\eta}(1))\right)\\
&=\frac{1}{n_{1k}n_{0k}}\sum_{i=1}^{n_{k}}D_{ik}(1-D_{ik})Cov\left(Y_{ik}(1),Y_{ik}(0)\middle|\textbf{D}_k,(\boldsymbol{\eta}(0),\boldsymbol{\eta}(1))\right)=0.
\end{align*}
The first equality holds because $Cov(Y_{ik}(d),Y_{jk}(d)|\textbf{D}_k,(\boldsymbol{\eta}(0), \boldsymbol{\eta}(1)))=Cov(Y_{ik}(d),Y_{jk}(d)|(\boldsymbol{\eta}(0), \boldsymbol{\eta}(1)))=0$ by Points \ref{hyp-point:indep-among-eps}, \ref{hyp-point:treatment-indep} and \ref{hyp-point:indep-eps-eta} of Assumption \ref{hyp:main_general}, the second equality holds because $D_{ik}(1-D_{ik})=0$. Therefore,
\begin{align}
&V\left(\widehat{ATE}_k\middle|(\boldsymbol{\eta}(0), \boldsymbol{\eta}(1))\right)\nonumber\\
&=\frac{1}{n_{0k}} S^2_{\E^*(Y_{ik}(0)),k} + \frac{1}{n_{1k}} S^2_{\E^*(Y_{ik}(1)),k} - \frac{1}{n_k} S^2_{\E^*(Y_{ik}(1))-\E^*(Y_{ik}(0)),k} \nonumber \\
&+\frac{1}{n_{1k}}\sum_{i=1}^{n_{1k}}\frac{1}{n_k}V\left(Y_{ik}(1)\middle|(\boldsymbol{\eta}(0),\boldsymbol{\eta}(1))\right)+ \frac{1}{n_{0k}}\sum_{i=1}^{n_{0k}}\frac{1}{n_k}V\left(Y_{ik}(0)\middle|(\boldsymbol{\eta}(0),\boldsymbol{\eta}(1))\right)\label{eq:noa1-cond-var}
\end{align}
Finally, by Lemma \ref{lem:conindep},
\begin{equation}
V\left(\widehat{ATE}\middle|(\boldsymbol{\eta}(0),\boldsymbol{\eta}(1))\right)=\frac{1}{K^2}\underset{k=1}{\overset{K}{\sum}}\left(\frac{n_k}{\overline{n}}\right)^2V\left(\widehat{ATE}_k|((\boldsymbol{\eta}(0), \boldsymbol{\eta}(1))\right).\label{eq:cond_var_as_wa_of_strata-level_cond_var}
\end{equation}

\subsubsection{Estimating an Upper Bound for the Conditional Variance of $\widehat{ATE}$}


%
%
\begin{align}
& \frac{1}{n_{1k}} \E^*\left[\frac{1}{n_{1k}-1}\underset{i=1}{\overset{n_k}{\sum}}D_{ik}\left(Y_{ik}(1)-\overline{Y}_{1k}\right)^2\right] \nonumber \\
&=\frac{1}{n_{1k}}\frac{1}{n_{1k}-1} \E^*\left[\underset{i=1}{\overset{n_k}{\sum}}D_{ik}Y_{ik}(1)^2-n_{1k}\overline{Y}_{1k}^2\right] \nonumber \\
&=\frac{1}{n_{1k}}\frac{1}{n_{1k}-1}\left[\underset{i=1}{\overset{n_k}{\sum}} \E^*(D_{ik}Y_{ik}(1)^2)-n_{1k} \E^*(\overline{Y}_{1k}^2)\right] \nonumber \\
&=\frac{1}{n_{1k}}\frac{1}{n_{1k}-1}\left[\underset{i=1}{\overset{n_k}{\sum}} \frac{n_{1k}}{n_k}\E^*(Y_{ik}(1)^2)-\frac{1}{n_{1k}}\underset{i=1}{\overset{n_k}{\sum}}\E^*(D_{ik}^2Y_{ik}^2(1))-\frac{1}{n_{1k}}\underset{i\neq j}{\sum\sum}\E^*(D_{ik}D_{jk}Y_{ik}(1)Y_{jk}(1))\right] \nonumber \\
&=\frac{1}{n_{1k}}\frac{1}{n_{1k}-1}\left[\underset{i=1}{\overset{n_k}{\sum}} \frac{n_{1k}}{n_k}\E^*(Y_{ik}(1)^2)-\frac{1}{n_{k}}\underset{i=1}{\overset{n_k}{\sum}}\E^*(Y_{ik}^2(1))-\frac{n_{1k}-1}{n_k(n_k-1)}\underset{i\neq j}{\sum\sum}\E^*(Y_{ik}(1))\E^*(Y_{jk}(1))\right] \nonumber \\
&=\frac{1}{n_{1k}}\frac{1}{n_{1k}-1}\left[\underset{i=1}{\overset{n_k}{\sum}} \frac{n_{1k}-1}{n_k}[V\left(Y_{ik}(1)\middle|(\boldsymbol{\eta}(0),\boldsymbol{\eta}(1))\right)+\E^*(Y_{ik}(1))^2]-\frac{n_{1k}-1}{n_k(n_k-1)}\underset{i\neq j}{\sum\sum}\E^*(Y_{ik}(1))\E^*(Y_{jk}(1))\right] \nonumber \\
&=\frac{1}{n_{1k}}\frac{1}{n_{1k}-1}\Bigg[\underset{i=1}{\overset{n_k}{\sum}} \frac{n_{1k}-1}{n_k}V\left(Y_{ik}(1)\middle|(\boldsymbol{\eta}(0),\boldsymbol{\eta}(1))\right) \nonumber \\
&+\frac{n_{1k}-1}{n_k}\underset{i=1}{\overset{n_k}{\sum}} \E^*(Y_{ik}(1))^2+\frac{n_{1k}-1}{n_k(n_k-1)}\underset{i}{\sum}\E^*(Y_{ik}(1))^2-\frac{n_{1k}-1}{n_k(n_k-1)}\underset{i,j}{\sum\sum}\E^*(Y_{ik}(1))\E^*(Y_{jk}(1))\Bigg] \nonumber \\
&=\frac{1}{n_{1k}}\left[\frac{1}{n_k}\underset{i=1}{\overset{n_k}{\sum}} V\left(Y_{ik}(1)\middle|(\boldsymbol{\eta}(0),\boldsymbol{\eta}(1))\right)\right]+ \frac{1}{n_{1k}}\frac{1}{n_k-1}\left[\underset{i=1}{\overset{n_k}{\sum}} \E^*(Y_{ik}(1))^2 - n_k \overline{\E^*(Y_{ik}(1))}^2\right] \nonumber \\
&=\frac{1}{n_{1k}} S^2_{\E^*(Y_{ik}(1)),k} +\frac{1}{n_{1k}}\sum_{i=1}^{n_{1k}}\frac{1}{n_k}V\left(Y_{ik}(1)\middle|(\boldsymbol{\eta}(0),\boldsymbol{\eta}(1))\right). \label{eq:noa1-upper_bound_cond_var_term1}
\end{align}
The third equality holds by Point  \ref{hyp-point:treatment-indep} of Assumption \ref{hyp:main_general} and by Equation \eqref{eq:ED}. The fourth equality holds because treatment is binary, by Point  \ref{hyp-point:treatment-indep} of Assumption \ref{hyp:main_general}, by Points \ref{hyp-point:indep-among-eps} and \ref{hyp-point:indep-eps-eta} of Assumption \ref{hyp:main_general}, and by  Equations \eqref{eq:ED} and \eqref{eq:cov}. The fifth equality holds by the definition of a conditional variance. The remaining equalities hold by algebraic manipulations. \medskip

Using similar arguments, one can show:
\begin{align}
&\E\left(\frac{1}{n_{0k}} \left[\frac{1}{n_{0k}-1}\underset{i=1}{\overset{n_k}{\sum}}\left(1-D_{ik}\right)\left(Y_{ik}(0)-\overline{Y}_{0k}\right)^2\right]\middle|(\boldsymbol{\eta}(0),\boldsymbol{\eta}(1))\right) \nonumber \\ =&\frac{1}{n_{0k}} S^2_{\E^*(Y_{ik}(0)),k}+ \frac{1}{n_{0k}}\sum_{i=1}^{n_{0k}}\frac{1}{n_k}V\left(Y_{ik}(0)\middle|(\boldsymbol{\eta}(0),\boldsymbol{\eta}(1))\right).\label{eq:noa1-upper_bound_cond_var_term2}
\end{align}

Equations \eqref{eq:noa1-cond-var}, \eqref{eq:noa1-upper_bound_cond_var_term1}, and  \eqref{eq:noa1-upper_bound_cond_var_term2} show that $V\left(\widehat{ATE}_k|(\boldsymbol{\eta}(0),\boldsymbol{\eta}(1))\right)\leq  \E^*\left(\widehat{V}_{rob}\left(\widehat{ATE}_k\right)\right)$.
Then, it follows from Equation \eqref{eq:cond_var_as_wa_of_strata-level_cond_var} and the definition of $\widehat{V}_{rob}\left(\widehat{ATE}_k\right)$ that
$$V\left(\widehat{ATE}|(\boldsymbol{\eta}(0),\boldsymbol{\eta}(1))\right)\leq  \E^*\left(\widehat{V}_{rob}\left(\widehat{ATE}\right)\right).$$

\subsubsection{Estimating an Upper Bound for the Variance of $\widehat{ATE}$ }

The proof in Section 6.3.3 does not make use of Assumption \ref{hyp:outcomes} so the result is already proven.

\end{document}